\documentclass[10pt,a4paper]{article}
\usepackage{myform,overcite}
\usepackage{tabularx,multirow,fancyhdr}
\usepackage{amssymb,mathrsfs}
\usepackage[footnotesize]{caption}
\usepackage{graphicx,epsfig,psfrag}
\usepackage{deleq}

\newcommand{\de}{\ensuremath{^{\circ}}} 
\newcommand{\integ}{\int\limits} 
\renewcommand{\d}{\ensuremath{\; \mathrm{d}}} 
\newcommand{\fig}[1]{Fig.~\ref{#1}} 
\newcommand{\eq}[1]{Eq.~(\ref{#1})} 
\newcommand{\vek}[1]{\ensuremath{\mathbf{#1}}} 
\providecommand{\url}[1]{\textcolor{blue}{\underline{#1}}}

\newcommand{\mm}{\ensuremath{\,\mathrm{mm}}}
\newcommand{\nm}{\ensuremath{\,\mathrm{nm}}}

\providecommand{\dash}{\ensuremath{^\prime}} %
\begin{document}
\setcounter{page}{5130}%
\twocolumn[%
\title{\bf \sf \huge Model for optical forward scattering by nonspherical raindrops}
\author{\sf \Large Oliver N. Ross and Stuart G. Bradley\footnotemark[2]}
\begin{abstract}
We describe a numerical model for the interaction of light with large raindrops using
realistic nonspherical drop shapes. We apply geometrical optics and a Monte Carlo technique to
perform ray traces through the drops. We solve the problem of diffraction independently by
approximating the drops with areaequivalent ellipsoids. Scattering patterns are obtained for
different polarizations of the incident light. They exhibit varying degrees of asymmetry and
depolarization that can be linked to the distortion and thus the size of the drops. The model
is extended to give a simplified long-path integration.
\end{abstract}
]
\footnotetext[2]{At the time of this research, O. N. Ross (o.ross@soton.ac.uk) was with Fachbereich Physik, Freie
Universit\"at Berlin, Berlin, Germany and is now with the School of Ocean and Earth Science, University of Southampton,
Southampton S014 3ZH, United Kingdom. S. G. Bradley was with the Department of Physics, University of Auckland, New
Zealand and is now with the School of Acoustics and Electronic Engineering, University of Salford, Salford M5 4WT,
United Kingdom.
\par Received 13 February 2002; accepted 6 May 2002.}
\section{Introduction}
\begin{figure}[b!]
\begin{center}
\epsfig{file=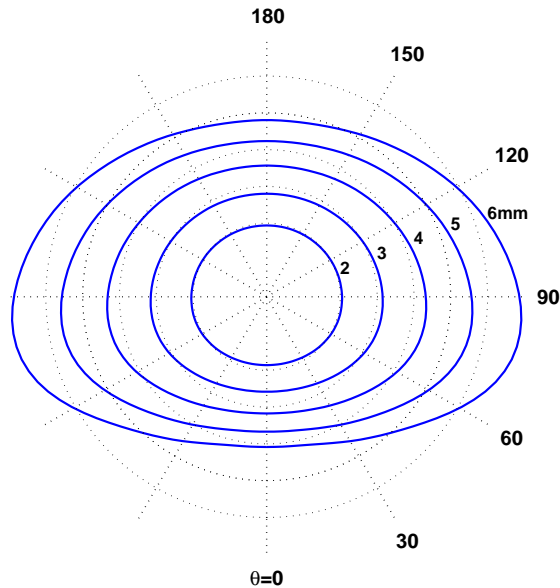,width=.9\columnwidth} 
\caption{Drop shapes for diameters $D=2$, 3, 4, 5 and 6\mm\ with dashed circles shown for
comparison and the angle $\theta$ in the coordinate system used.} \label{fig1}
\end{center}
\end{figure}

Large raindrops are deformed from spheres because of aerodynamic pressure as they fall at terminal
velocity.\cite{lenard,magono,pruppacher} Beard and Chuang's\cite{beard} (and Chuang and \linebreak
Beard's\cite{chuang}) model results for the flattening at the base of the drops are used in this study. Drop shapes are
specified by
\begin{equation}
r(\theta )=a[1+\sum_{n=0}^{10}c_{n}\cos (n\theta )]  \label{fit}
\end{equation}
where $a$ is the radius of the undistorted sphere, located at the center of mass of the drop;
$\theta$ is the polar elevation with $\theta = 0$\de\ pointing vertically downward
(\fig{fig1}) and $c_n$ are shape co-efficients given in Table~\ref{tab:5}. Most previous
research is restricted to spherical drops\cite{glantschnig,kazovsky2,lock} or ellipsoid
particles.\cite{stamaskos} Macke and Großklaus\cite{macke} used the same distorted drop shapes
as the present study for backscattering applicable to lidar measurements of rainfall
intensities. In the current study we are interested in using drop distortion in a
forward-scatter mode so as to measure drop size and, for long paths, rainfall intensity. A
complete description of the model, including its application to a long path integration, can
be found in Ross\cite{ross}.
\section{The Ray Trace Method} \label{carlo} \label{sec:GOA}
We used a Monte Carlo approach in which a large number of photons are traced through the drop
using geometrical optics. Diffraction is accounted for analytically. For our simulation we
chose a wavelength of $\lambda = 650\nm$, and thus the size parameter $x=2\pi a/\lambda$ is in
the range $ 5\cdot 10^{3} \lesssim x \lesssim 4\cdot10^{4}$ for the drop sizes in
Table~\ref{tab:5}. Glantschnig and Chen\cite{glantschnig} found that geometrical optics
produced good results in comparison with the rigorous Mie theory for $x \ge 20$ and scattering
angles $\le 60\de$. Macke {\it et al.}\/ \cite{macke2} found this condition to be $x \ge 60$.
In a different study,\cite{shah} computations were carried out over a wide range of size
parameters. The results showed that the deviation from Mie theory is less than 1\% for size
parameters $x\approx 10^4$ and $m=1.33$, confirming the validity of our geometrical optics
approach.
\subsection{Forward Scattering}
Table~\ref{tab:4}, adapted from van de Hulst,\cite{hulst} gives the forward- and backward-scattered intensities for
spherical scatterers and polarizations perpendicular and parallel to the scattering plane (polarization 1 and 2,
respectively). Over 91\% of light of polarization 1 and more than 97\% of polarization 2 is forward scattered. As can
be seen from Table~\ref{tab:4}, upward of 99.5\% of the total forward-scattered light for both polarizations emerges
from the first interface after simple reflection ($p=0$) and from the second interface after twofold refraction ($p =
1$). The fraction of forward-scattered intensity in the $p=0$ and $p=1$ rays increases for more distorted drops (see
Section 4), and so it is sufficient to consider only the contributions of these rays to forward scattering.
\subsection{The Procedure} \label{sec:proc}
\begin{figure}[t!]
\begin{center}
\epsfig{file=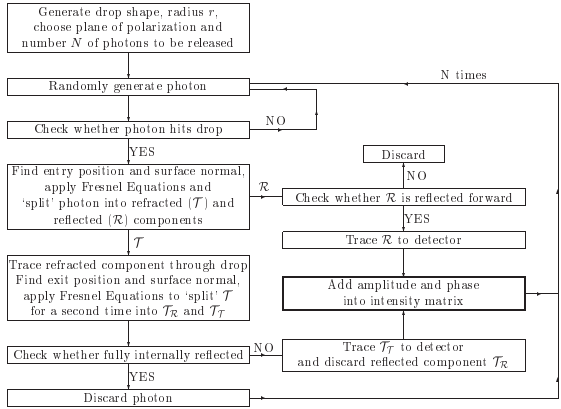,width=.9\columnwidth}
\end{center}
\caption{Flowchart for the ray trace.} \label{flowchrt}
\end{figure}

\fig{flowchrt} shows the flow diagram for the ray trace model. The model is run with $N=10^8$
photons. Runs for different polarizations of the incident light use the same set of random
photons to exclude statistical variations as a possible cause for differences in the
\begin{table*}
\caption{Shape co-efficients for cosine distortion fit (\eq{fit}) for drop radii between 0.5
and 4.5\mm\ from Chuang and Beard\cite{chuang}.} \label{tab:5}
\begin{center}
\footnotesize
\newcolumntype{C}{>{\centering\arraybackslash}X}%
\newcolumntype{R}{>{\raggedleft\arraybackslash}X}%
\newcolumntype{L}{>{\raggedright\arraybackslash}X}%
\newcolumntype{S}{>{\setlength{\hsize}{1.22\hsize}}C}%
\newcolumntype{V}{>{\setlength{\hsize}{0.98\hsize}}C}%
\begin{tabularx}{\textwidth}{S V V V V V V V V V V V}
\hline \\[-6pt]
\multirow{2}{28mm}{\shortstack{a \\[1pt] [mm]}}
& \multicolumn{11}{c}{Shape co-efficients ($c_n\cdot 10^{4}$) for $n=$}
\\[4pt] \cline{2-12} \\[-8pt]
  & 0 & 1 & 2 & 3 & 4 & 5 & 6 & 7 & 8 & 9 & 10 \\ \hline \hline
\end{tabularx}
\begin{tabularx}{\textwidth}{L R R R R R R R R R R R}
0.5 & -28 & -30 & -83 & -22 & -3 & 2 & 1 & 0 & 0 & 0 & 0 \\
0.75 & -72 & -70 & -210 & -57 & -6 & 7 & 3 & 0 & -1 & 0 & 1 \\
1.0 & -134 & -118 & -385 & -100 & -5 & 17 & 6 & -1 & -3 & -1 & 1 \\
1.25 & -211 & -180 & -592 & -147 & 4 & 32 & 10 & -3 & -5 & -1 & 2 \\
1.5 & -297 & -247 & -816 & -188 & 24 & 52 & 13 & -8 & -8 & -1 & 4 \\
1.75 & -388 & -309 & -1042 & -221 & 53 & 75 & 15 & -15 & -12 & 0 & 7 \\
2.0 & -481 & -359 & -1263 & -244 & 91 & 99 & 15 & -25 & -16 & 2 & 10 \\
2.25 & -573 & -401 & -1474 & -255 & 137 & 121 & 11 & -36 & -19 & 6 & 13 \\
2.5 & -665 & -435 & -1674 & -258 & 187 & 141 & 4 & -48 & -21 & 11 & 17 \\
2.75 & -755 & -465 & -1863 & -251 & 242 & 157 & -7 & -61 & -21 & 17 & 21 \\
3.0 & -843 & -472 & -2040 & -240 & 299 & 168 & -21 & -73 & -20 & 25 & 24 \\
3.25 & -930 & -487 & -2207 & -222 & 358 & 175 & -37 & -84 & -16 & 34 & 27 \\
3.5 & -1014 & -492 & -2364 & -199 & 419 & 178 & -56 & -93 & -12 & 43 & 30 \\
4.0 & -1187 & -482 & -2650 & -148 & 543 & 171 & -100 & -107 & 2 & 64 & 32 \\
4.5 & -1328 & -403 & -2889 & -106 & 662 & 153 & -146 & -111 & 18 & 81 & 31 \\
\hline
\end{tabularx}
\end{center}
\end{table*}
scattering patterns. The photons propagate parallel to the $x$ axis from negative $x$ toward
\begin{table}[b!]
\caption{Separation into forward and backward scattering for polarisations 1 (first number in
each column) and 2.} \label{tab:4}
\begin{center}
\newcolumntype{C}{>{\centering\arraybackslash}X}%
\newcolumntype{R}{>{\raggedleft\arraybackslash}X}%
\newcolumntype{S}{>{\setlength{\hsize}{0.58\hsize}}C}%
\newcolumntype{V}{>{\setlength{\hsize}{1.14\hsize}}R}%
{\small
\begin{tabularx}{\columnwidth}{SS VV VV VV}
\hline \\[-5pt] \multicolumn{2}{c}{Ray} & \multicolumn{2}{c}{Forward} &
\multicolumn{2}{c}{Backward} & \multicolumn{2}{c}{Total} \\[2mm] \hline
\multicolumn{2}{c}{$p=0$} & 857 & 248 & 163 & 59 & 1020 & 307 \\
\multicolumn{2}{c}{$p=1$} & 8217 & 9456 & 0 & 0 & 8217 & 9456 \\
\multicolumn{2}{c}{combined} & 9074 & 9704 & 163 & 59 & 9237 & 9763 \\
\multicolumn{2}{c}{$p\ge2$} & 44 & 15 & 719 & 222 & 763 & 237 \\[5pt]
\multicolumn{2}{c}{All $p$} & 9118 & 9719 & 882 & 281 & 10,000 & 10,000 \\ \hline
\end{tabularx}
}
\end{center}
\end{table}
the drop, and a point \vek{P} on the drop's surface can be represented in Cartesian
coordinates by
\begin{equation}
\vek{P}=r(\theta)\left[
\begin{array}{c}
-\sin{\phi}\sin{\theta} \\
\cos{\phi}\sin{\theta}    \\
-\cos{\theta}
\end{array}
\right] \label{surf}
\end{equation}
where $\phi$ is the azimuth angle, increasing clockwise from the $y$-axis. The surface normal
at \vek{P} is
\begin{equation}
\vek{N} = \left(
\frac{\partial{\vek{P}}}{\partial\theta} \times
\frac{\partial{\vek{P}}}{\partial\phi} \right)\, .
\end{equation}
Using Snell's law and the Fresnel equations, we can now trace the photons through the drop.
The detector coordinates ($\Theta$,$\Phi$) are defined similarly to ($\theta$,$\phi$), and
arriving photons are resolved into angular bins of 1\de\ resolution in $\Theta$ and $\Phi$.
Each photon generated is assigned a weight of one. The intensity fractions for $p=0$ and $p=1$
are stored in an intensity matrix \vek{I} with $\Theta$ and $\Phi$ as row and column indices,
respectively.
\section{Diffraction} \label{difsec}
Analytic solutions for diffraction from distorted drop shapes do not exist. Here, drop shapes
are approximated by area-equivalent ellipsoids, for which the diffraction pattern is known.
This approximation is verified when the exact diffraction problem is numerically solved for
one of the larger drops. Results showed negligible differences at the angular resolution used.
\begin{figure}[b!]
\begin{center}
\epsfig{file=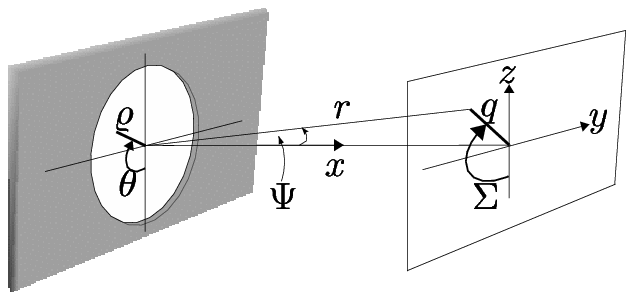,width=.9\columnwidth,clip=true}
\end{center}
\caption{Co-ordinate system used for the diffraction integral in \eq{diff}.}
\label{fig:difcoord}
\end{figure}
Given the range of size parameters in this study, diffraction scatters energy equal to
refraction and reflection combined,\cite{bohren,hulst} i.e., the extinction efficiency is 2.
\begin{figure}[t!]
\begin{center}
\epsfig{file=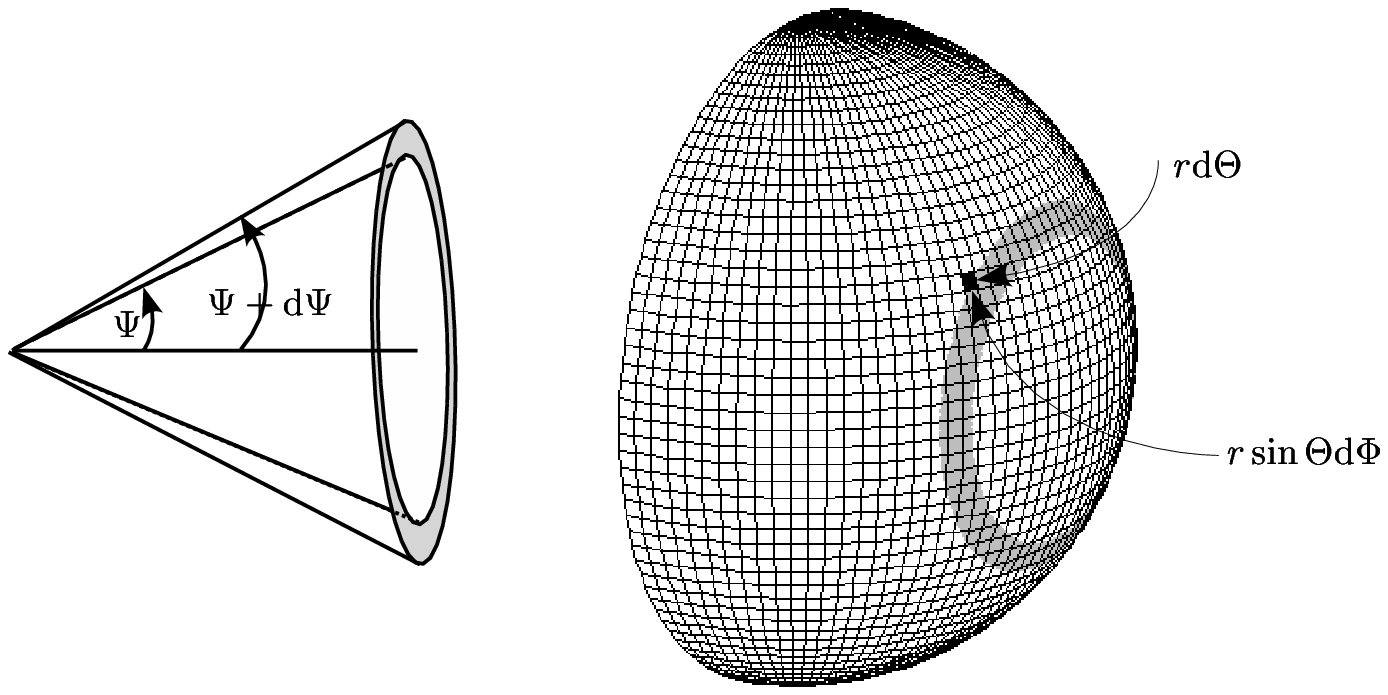,width=\columnwidth,clip=true}  
\end{center}
\caption{The energy ring (left) and its projection onto the spherical detector.}
\label{diffract}
\end{figure}
\subsection{Diffraction from Spheres}
The framework for the elliptical approximation is based on diffraction from a spherical
aperture\cite{hecht,born}:
\begin{eqnarray}
\mathscr{E}(q,\Sigma) & = & \frac{{\mathcal{E}}_A \ \exp{[i(\omega t-kr)]}}{r} \
\integ_{\varrho=0}^a \integ_{\theta=0}^{2\pi} \ \exp{[i(k \varrho q /r)} \nonumber \\
& & \times \cos(\theta-\Sigma)] \ \varrho \ d\varrho \ d\theta \, ,  \label{diff}
\end{eqnarray}
where $\mathscr{E}$ is the electric field amplitude, giving intensity $I\propto|
\mathscr{E}|\/^2 $. ${\mathcal{E}}_A$ denotes the source strength per unit area of the
aperture; and $r$ the distance between its center and the detector (\fig{fig:difcoord}). The
solution is
\begin{equation}
I \propto  \left[ \frac{\mathscr{J}_1(kaq/r)}{kaq/r} \right] ^2 = \left[
\frac{\mathscr{J}_1(ka\sin\Psi)}{ka\sin\Psi} \right] ^2 \, ,\label{circsol}
\end{equation}
where $a$ is the radius of the aperture, $k$ is the wave number 2$\pi/\lambda$, and
$\mathscr{J}_1$ is the Bessel function of the first kind and order one. The solution does not
depend on the angle $\Sigma$ due to the circular symmetry. By integrating
relation~\ref{circsol}, we write the fraction of the total intensity contained within a cone
of angle $\Psi$ as
\begin{equation}
L(\Psi)=1-\mathscr{J}_0^2(ak\sin(\Psi))-\mathscr{J}_1^2(ak\sin(\Psi)) \, , \label{ener}
\end{equation}
where $\mathscr{J}_0$ is the Bessel function of order zero\cite{towne}.
\par
If the energy flux through a ring element of area $2\pi r^2\sin\Psi\d\Psi$ (\fig{diffract}) is
$L\dash(\Psi) \d\Psi$, then the fractional flux through a detector element of angular
dimensions (d$\Theta$,d$\Phi$) and area $r^2\sin\Theta\d\Theta\d\Phi$ is
\begin{eqnarray}
L\dash(\Psi)\frac{\sin\Theta\d\Theta\d\Phi}{2\pi\sin\Psi}
\end{eqnarray}
which can be expressed in terms of detector coordinates because $\cos\Psi=\sin\Theta\sin\Phi$.
After this substitution and further simplifications, the final result is
\begin{eqnarray}
\tilde{I}_{\Theta\Phi}  =  \frac{1}{2\pi}\integ_{\Delta\Theta}\integ_{\Delta\Phi}
                             \frac{2\mathscr{J}_1^2(ak\xi) \sin^2\Theta \sin\Phi}%
                             {\xi^2}\d\Theta\d\Phi  \label{boxint}
\end{eqnarray}
where $\xi$ stands for $\left[ \cos^2{\Theta}+\cos^2{\Phi} \sin^2{\Theta} \right]^{1/2} $. The
intervals $\Delta\Theta$ and $\Delta\Phi$ are appropriately chosen to cover the size of the
bin. By analogy with the intensity matrix introduced in Subsection 2.B, this yields the
$181\times 181$ intensity matrix $\tilde{\vek{I}}$ for diffraction, which we find numerically.
\subsection{Approximating the Drop Shapes with Ellipsoids}
\begin{figure}[t!]
\begin{center}
\epsfig{file=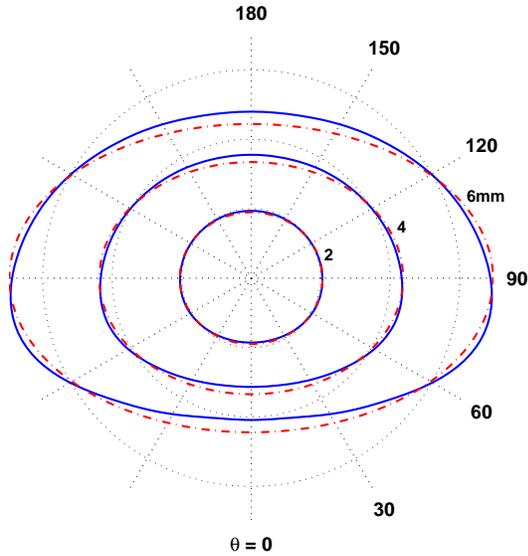,width=.85\columnwidth,clip=true}
\end{center}
\caption{Comparison between the ellipses used (dash-dot line) and the actual drop shapes for
drop diameters $D=2$, 4 and 6\mm.} \label{fig:comp}
\end{figure}

The Fraunhofer diffraction integral for the distorted drops has no analytic solution as $\varrho$ is a function of
$\theta$. Hence we approximate the drops with areaequivalent ellipsoids. The cross-sectional area of the drops can be
calculated from
\begin{eqnarray}
A & = & \frac{1}{2}\oint_C r^2(\theta)\d\theta =
a^2\integ_0^{\pi}\left[1+\sum_{n=0}^{10}c_n\cos(n\theta)\right]^2\d\theta
\nonumber \\
  & = & \pi a^2\left[1+2c_0+c_0^2+\sum_{n=1}^{10}\frac{c_n^2}{2}\right] \label{area}
\end{eqnarray}
with the co-efficients $c_n$ from Table \ref{tab:5}. The discrepancy between the approximation
and the actual drop shapes is the largest for the bigger drops and vanishes as the size (and
the flattening at the base) of the drop decreases. \fig{fig:comp} shows some examples.
\par
Extending the circular aperture in one direction by a constant factor $\mu$ will cause the
diffraction pattern to contract in that direction by the same factor. Because of the $\mu$
times larger area of the aperture, the intensity is $\mu^2$ times the original intensity at
each point mapped from the original pattern.\cite{francon,born} Hence we can calculate the
contribution to each bin from diffraction using elliptic obstacles from the results of a
circular aperture.
\subsection{Verifying the Elliptic Approximation}
In this subsection the diffraction integral is evaluated numerically for one exact drop shape.
The results are used to validate the elliptic approximation. The constant radius $\varrho$ in
\eq{diff} is replaced by the cosine distortion fit from \eq{fit}, and the variable of
integration is changed from $\varrho$ to $a$, the radius of the undistorted sphere. \eq{diff}
becomes
\begin{eqnarray}
\lefteqn{\mathscr{E}(q,\Sigma) =  \underbrace{\frac{\mathcal{E}_A \: \exp{[i(\omega t-kr)]}}{r}}_{C} \integ_{a=0}^{R}
\integ_{\theta=0}^{2\pi} } \nonumber \\
& \times & \exp \left[i(kqa/r) \, \cos(\theta-\Sigma) \left(1+\sum_{n=0}^{10} c_n \, \cos(n \theta)\right)\right]
\nonumber \\
& \times & a \, \left(1+\sum_{n=0}^{10} c_n \, \cos(n \theta)\right)^2 \d a \d\theta \, . \label{start}
\end{eqnarray}
To simplify \eq{start}, the following substitutions can be applied
\begin{eqnarray}
\chi =  \frac{kq}{r} \cos(\theta - \Sigma), \xi = \left(1+\sum_{n=0}^{10} c_n \cos(n
\theta)\right)  \label{subs}
\end{eqnarray}
\vspace*{-22pt}
\begin{eqnarray}
\lefteqn{\Rightarrow \mathscr{E}(q,\Sigma) = C \integ_{a=0}^{R} \integ_{\theta=0}^{2\pi}
                       \exp{(i \chi \xi a)} \: a \, \xi^2 \d a \d\theta } \\
&=&  - \, C \integ_{\theta=0}^{2\pi}
                         \left[\frac{\exp{(i \chi \xi \, a)}}{\chi^2 \xi^2} \:
                        (i \chi \xi \, a -1) \right]_{a=0}^R \, \xi^2 \d\theta \\
&=& - \, C \integ_{\theta=0}^{2\pi}
                        \frac{\exp{(i \chi \xi R)} \, (i \chi \xi R -1) +1}{\chi^2}
                        \d\theta  \, . \hspace{0mm} \label{simpl}
\end{eqnarray}
The integrand in \eq{simpl} is resolved into its real and imaginary components
\begin{eqnarray}
\mathscr{E}(q,\Sigma) = & C & \left[ \;\integ_{\theta=0}^{2\pi}
                \frac{\chi \xi R \sin(\chi \xi R)+\cos(\chi \xi
                R)-1}{\chi^2} \d\theta \right. \nonumber \\
              & + & \left. i \integ_{\theta=0}^{2\pi}
               \frac{\sin(\chi \xi R)-\chi \xi R\cos(\chi \xi
                R)}{\chi^2} \d\theta \right] \label{end}
\end{eqnarray}
which can be evaluated numerically for a given pair of detector coordinates (q,$\Sigma$). To
return to the original detector coordinates $(\Theta,\Phi)$ it can be shown that
\begin{equation}
\frac{q}{r}\cos(\theta-\Sigma) = \cos\theta\cos\Theta-\sin\theta\cos\Phi\, . \label{equiv}
\end{equation}
Making the replacements in \eq{start} and (\ref{subs}) leads to the same integral as in
\eq{simpl} except that $\chi$ now represents $k$ times the right-hand side of \eq{equiv}
instead of the left-hand side.
\par
\begin{figure}[t!]
\begin{center}
\epsfig{file=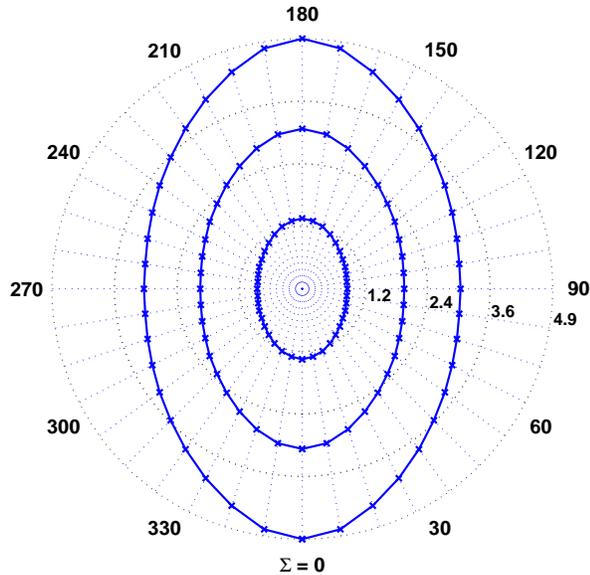,width=.95\columnwidth,draft=false}
\end{center}
\caption{Location of the first three minima in the diffraction pattern from a drop with radius
$a=3$\mm.} \label{elipdif}
\end{figure}

Because $I\propto |\mathscr{E}|^2$, only the real part of \eq{end} needs to be solved.
\fig{elipdif} shows the results obtained for a distorted drop with radius $a=3\mm$. The
locations of the first three minima exhibit a clear elliptic symmetry. The dotted lines
emerging radially from the center represent the lines along which the integrations were
carried out. The dotted circles with numbers next to them give the lines of constant $\Psi$ as
percentages of a degree. The resolution for the integration was 10,000 points/deg, which gives
approximately 500 points from the center of \fig{elipdif} to the outermost circle.
\par
A continuous plot of the field amplitudes $\mathscr{E}(q,0^{\circ})$ and
\begin{figure}[b!]
\begin{center}
\epsfig{file=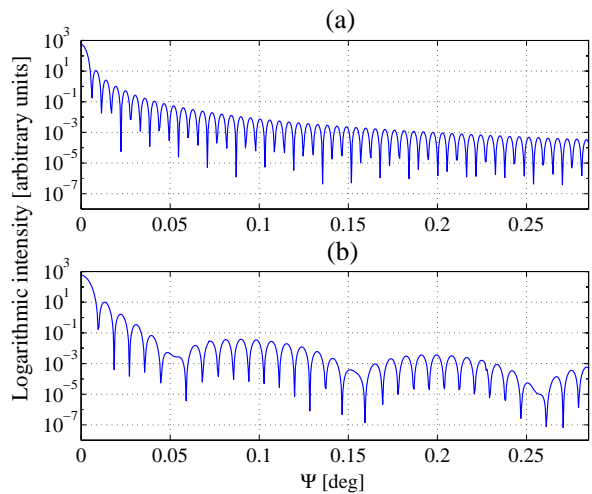,width=.95\columnwidth}
\end{center}
\caption{Logarithmic intensity distribution for the directions (a) $\Sigma=90\de$ and (b)
$\Sigma=0\de$.} \label{fig:logint}
\end{figure}
$\mathscr{E}(q,90^{\circ})$ is shown in \fig{fig:logint}. The eccentricity of the pattern is
obvious. The first minimum in the $\Sigma=90$\de\ direction is considerably closer to the
center, indicating the contraction in the horizontal. A similar comparison between
$\mathscr{E}(q,0^{\circ})$ and $\mathscr{E}(q,180^{\circ})$ (not shown) yields two identical
curves within the margin of error, confirming the symmetry of the pattern about the
horizontal. An interesting feature is the somewhat elongated sixth maximum in the
$\Sigma=0\de$ direction. This feature returns in a similar shape with every 11th maximum.
\par
A more quantitative account of the possible flattening in the pattern for the 3-mm drop can be
obtained from
$\aleph(q)=(\mathscr{E}(q,270\de)-\mathscr{E}(q,90^{\circ}))/\mathscr{E}(q,90\de)$, which
gives a measure for the relative deviation from horizontal symmetry. Values do not exceed
0.4\% and are only notably over 0.1\% within 0.1\de\ of the central forward direction. We
believe that this fully justifies the elliptic approximation, especially given the applied
angular resolution of 1\de\ and also considering that the distortion decreases for smaller
drops. The same computations can be carried out along the $\Sigma=0$ and $\Sigma=180$\de\
directions. Values in the absolute deviation do not exceed $4\cdot 10^{-6}$, which is the
region of error applied during the numerical integration and can thus be neglected.

\section{Results} \label{sec:res}

The results from the computer model are presented separately for the ray trace (no
diffraction) and for scattering including diffraction. The scattering patterns are displayed
by a sinusoidal projection method (which means that the meridians appear as sine functions),
hence the scale is preserved only on the central meridian and along the lines of latitude.
This projection method also renders the region around the central forward direction without
distortions.
%
\subsection{Scattering from Single Drops -- excluding Diffraction} \label{sec:res1}
%
The phase is omitted because the path lengths of the photons arriving at a particular bin can
differ by several tens of wavelengths. These effects cancel out for a large number of photons
and do not contribute any valuable information to the pattern.
\par
\begin{figure*}[t!]
\begin{minipage}[t]{0.95\columnwidth}
\footnotesize \psfrag{a}[c][c]{(a)}
\epsfig{file=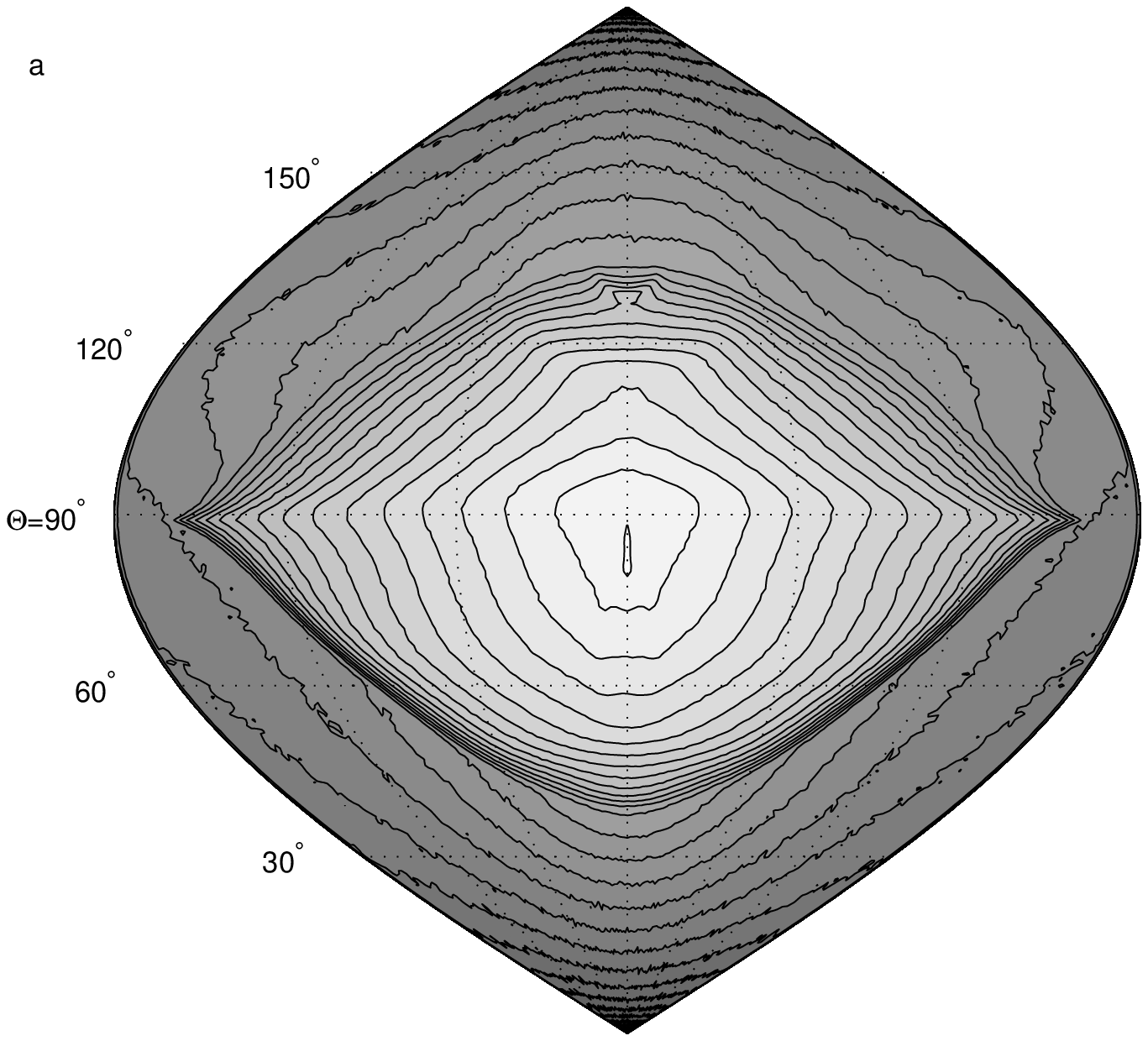,width=.95\columnwidth,draft=false}  
\end{minipage}
\hfill
\begin{minipage}[t]{0.95\columnwidth}
\footnotesize \psfrag{b}[c][c]{(b)}
\epsfig{file=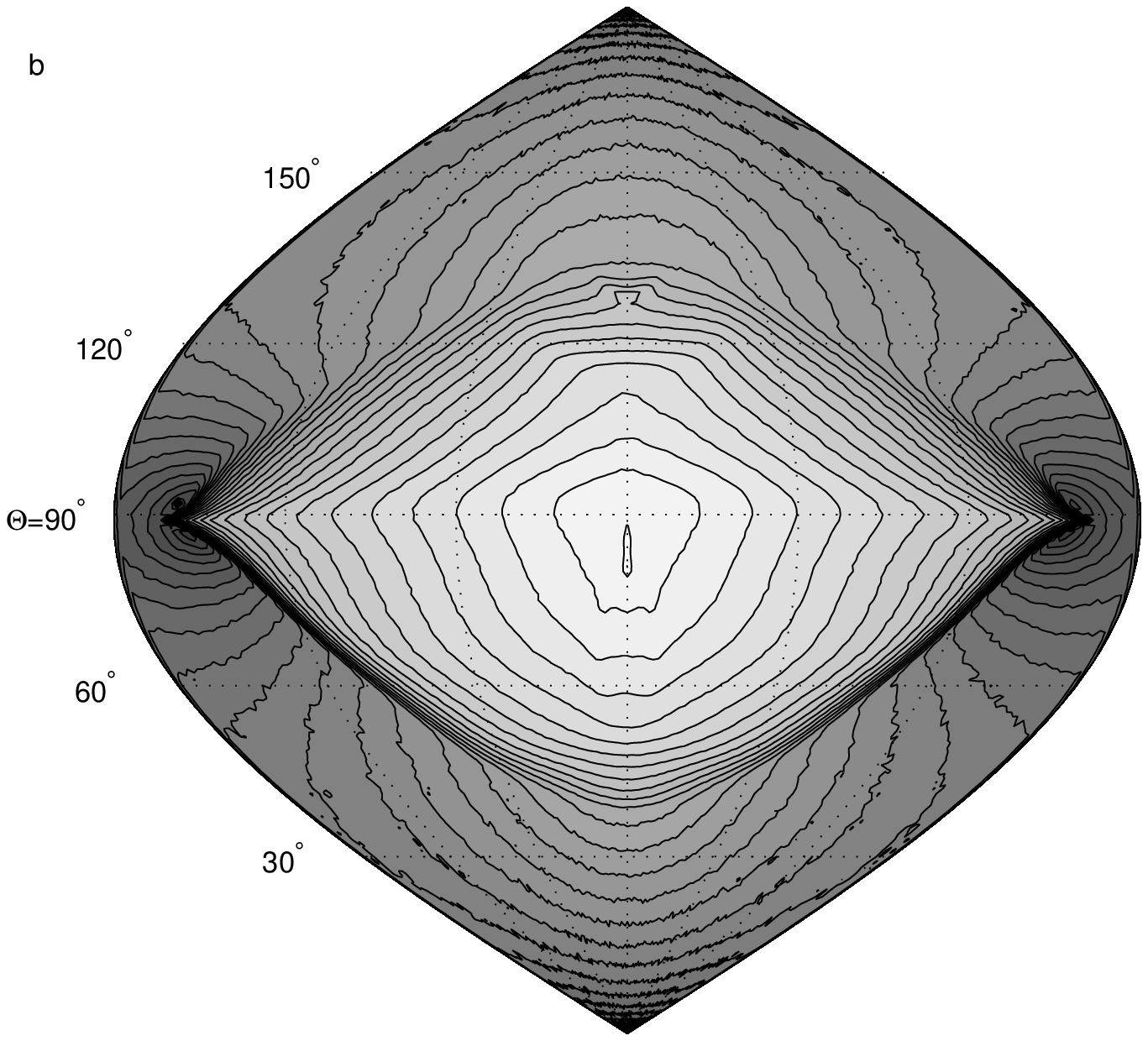,width=.95\columnwidth,draft=false}  
\end{minipage}
\\[15pt]
\begin{minipage}[t]{0.95\columnwidth}
\footnotesize \psfrag{c}[c][c]{(c)}
\epsfig{file=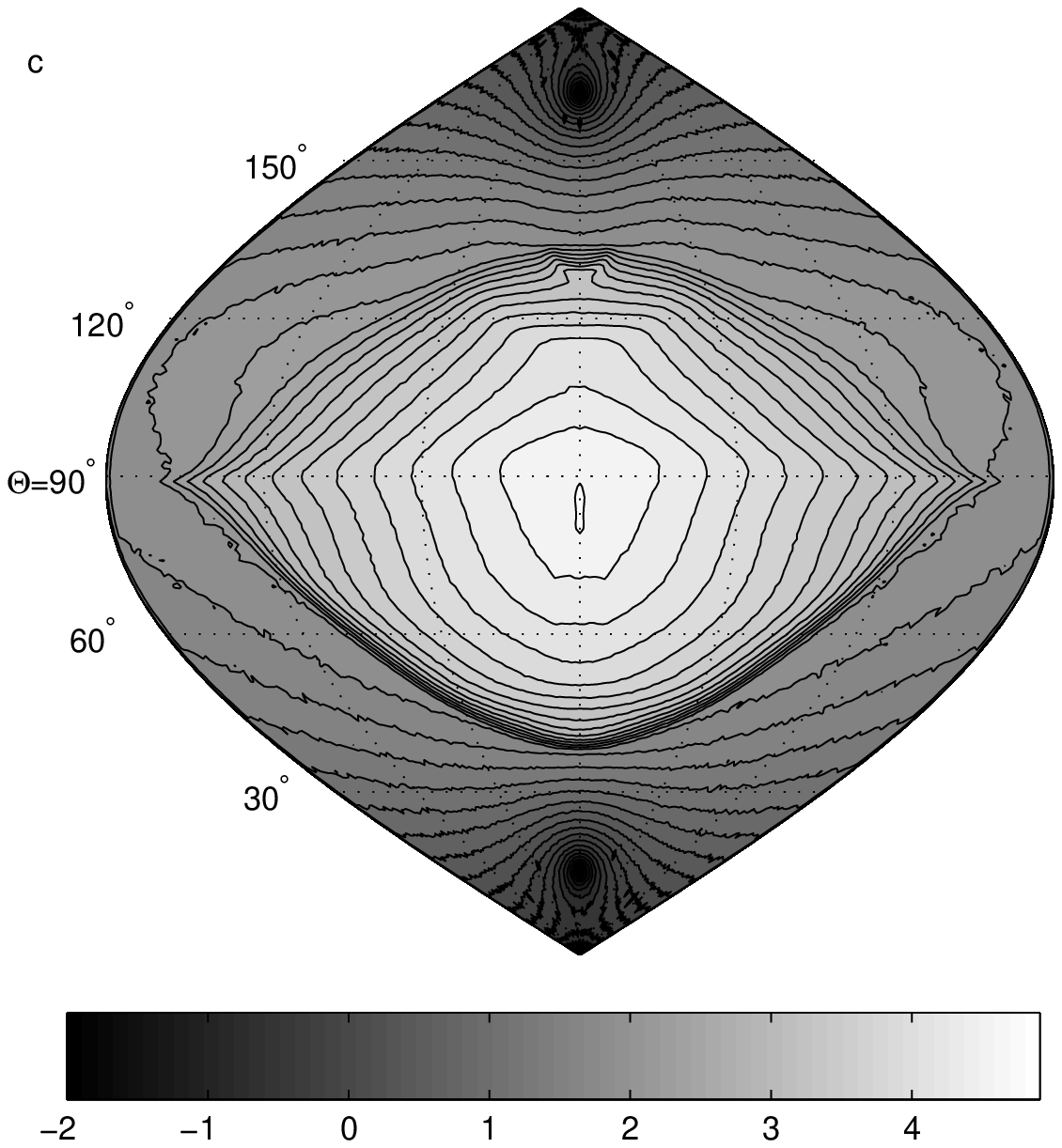,width=.95\columnwidth,draft=false}  
\end{minipage}
\hfill
\begin{minipage}[t]{0.95\columnwidth}
\footnotesize \psfrag{d}[c][c]{(d)}
\epsfig{file=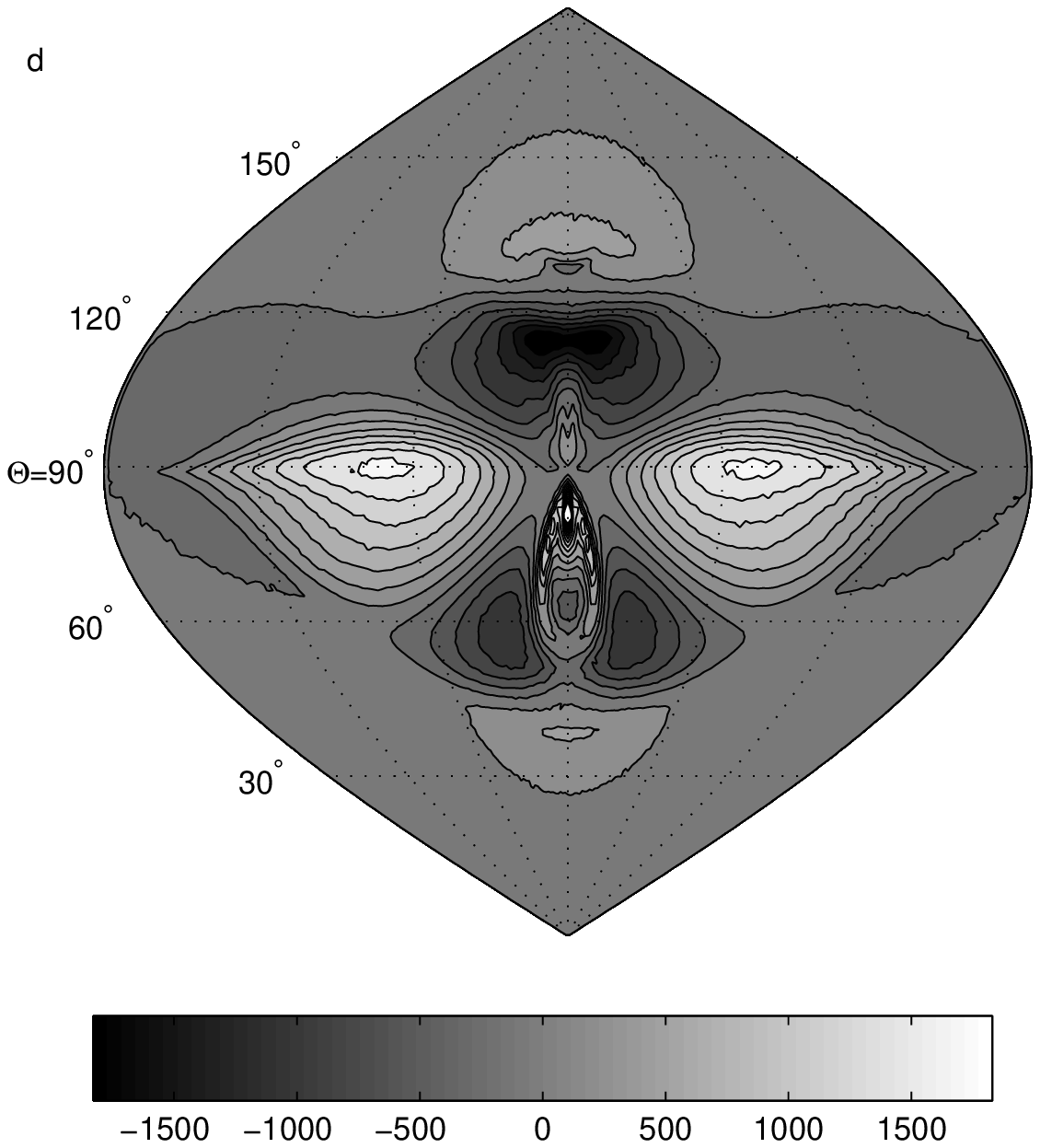,width=.95\columnwidth,draft=false}  
\end{minipage}
\caption{Contour plots of the intensity distribution obtained from a 3\mm\ drop, for: (a)
unpolarised light, (b) $y$- and (c) $z$-polarised light. A plot of the difference in the
obtained intensities for $y$-polarised and $z$-polarised light (non-logarithmic) is shown in
(d). Note that the gray scale below (c) applies for (a) to (c) while (d) being the only
non-logarithmic plot has a scale of its own.} \label{fig:cont2}
\end{figure*}
\begin{figure*}[t!]
\small
\begin{minipage}[t]{0.95\columnwidth}
\psfrag{a}[c][c]{(a)}
\epsfig{file=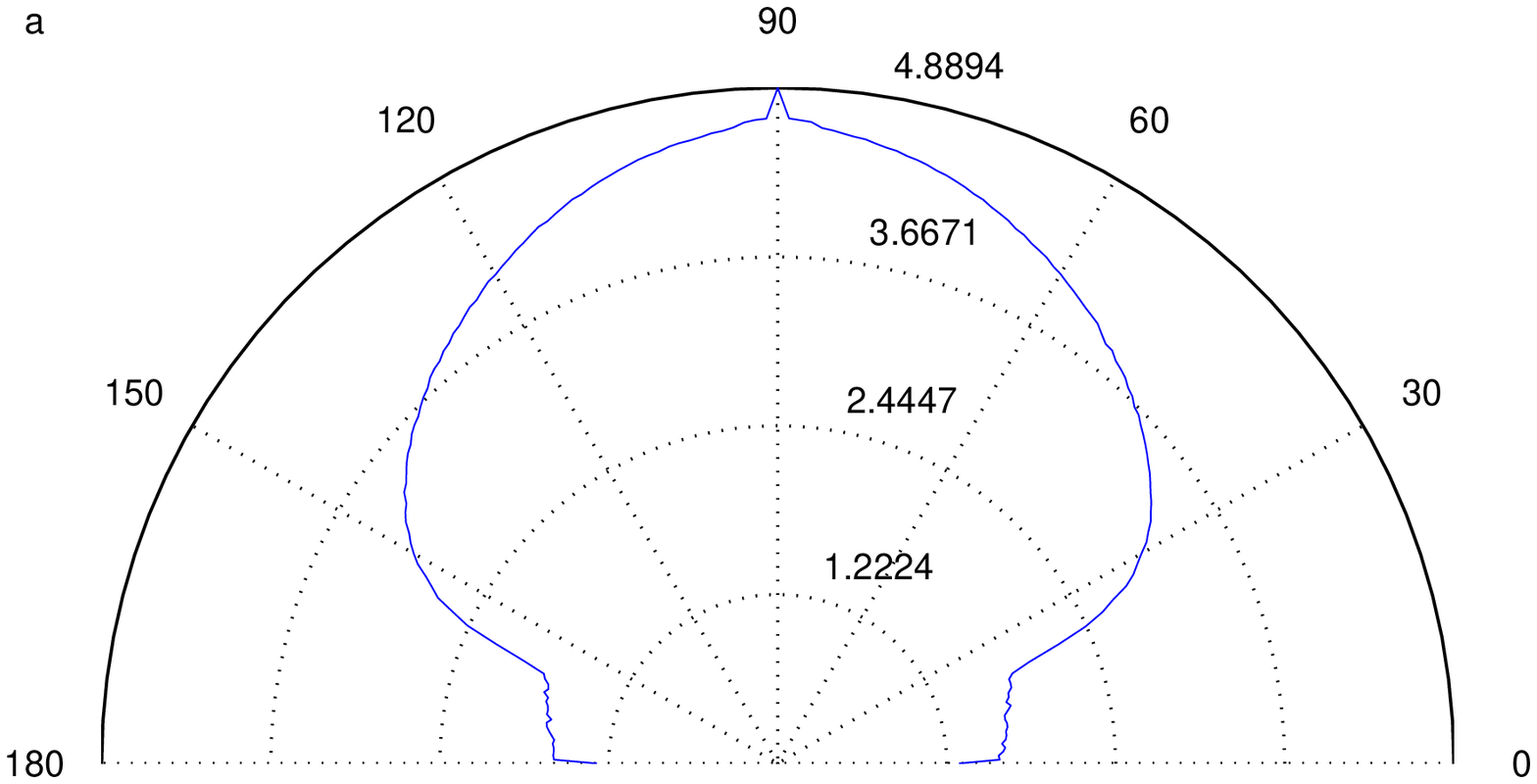,width=0.95\columnwidth,draft=false}  
\end{minipage}
\hfill
\begin{minipage}[t]{0.95\columnwidth}
\psfrag{b}[c][c]{(b)}
\epsfig{file=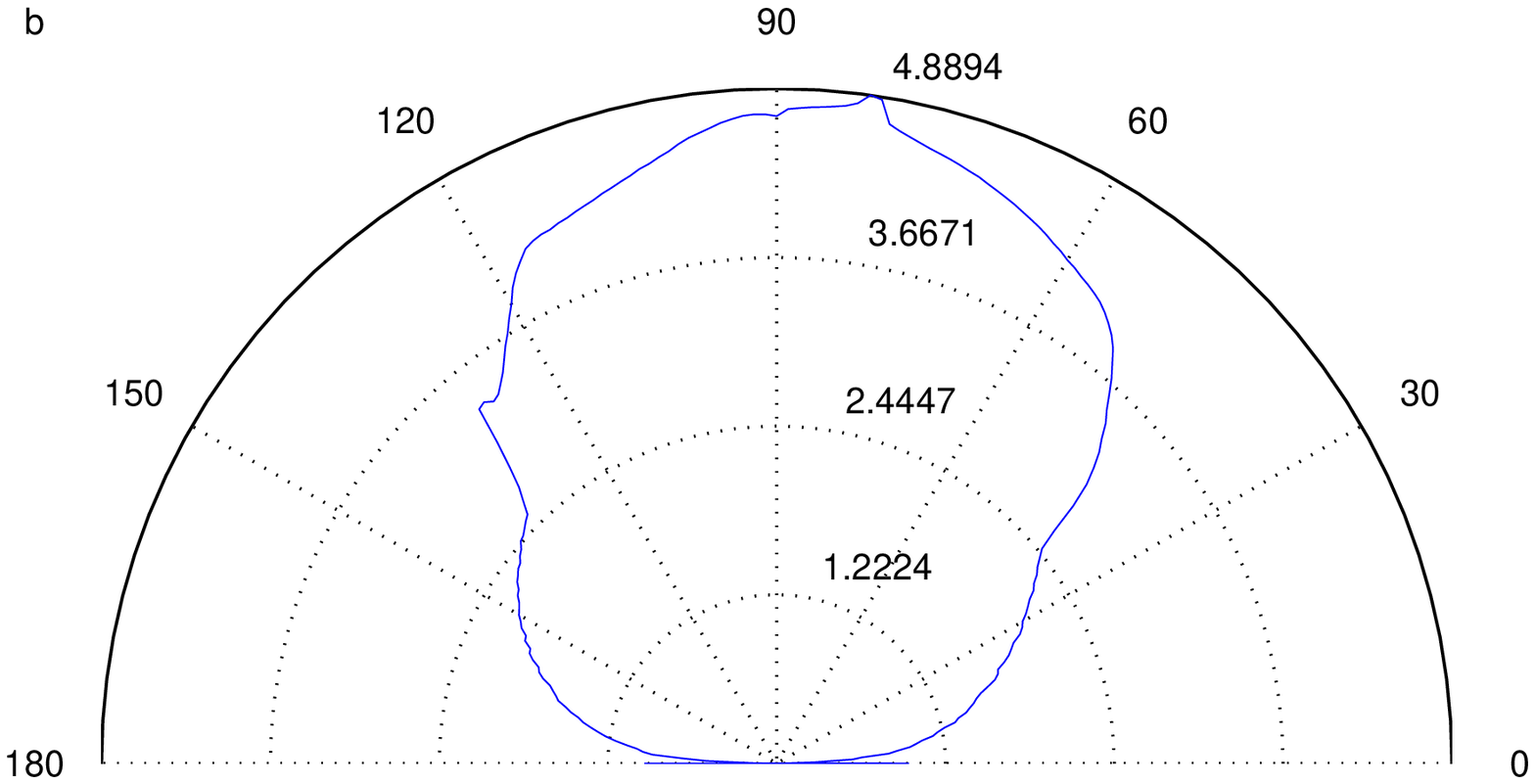,width=0.95\columnwidth,draft=false}  
\end{minipage}
\\[5pt]
\begin{minipage}[t]{0.95\columnwidth}
\psfrag{a}[c][c]{(c)}
\epsfig{file=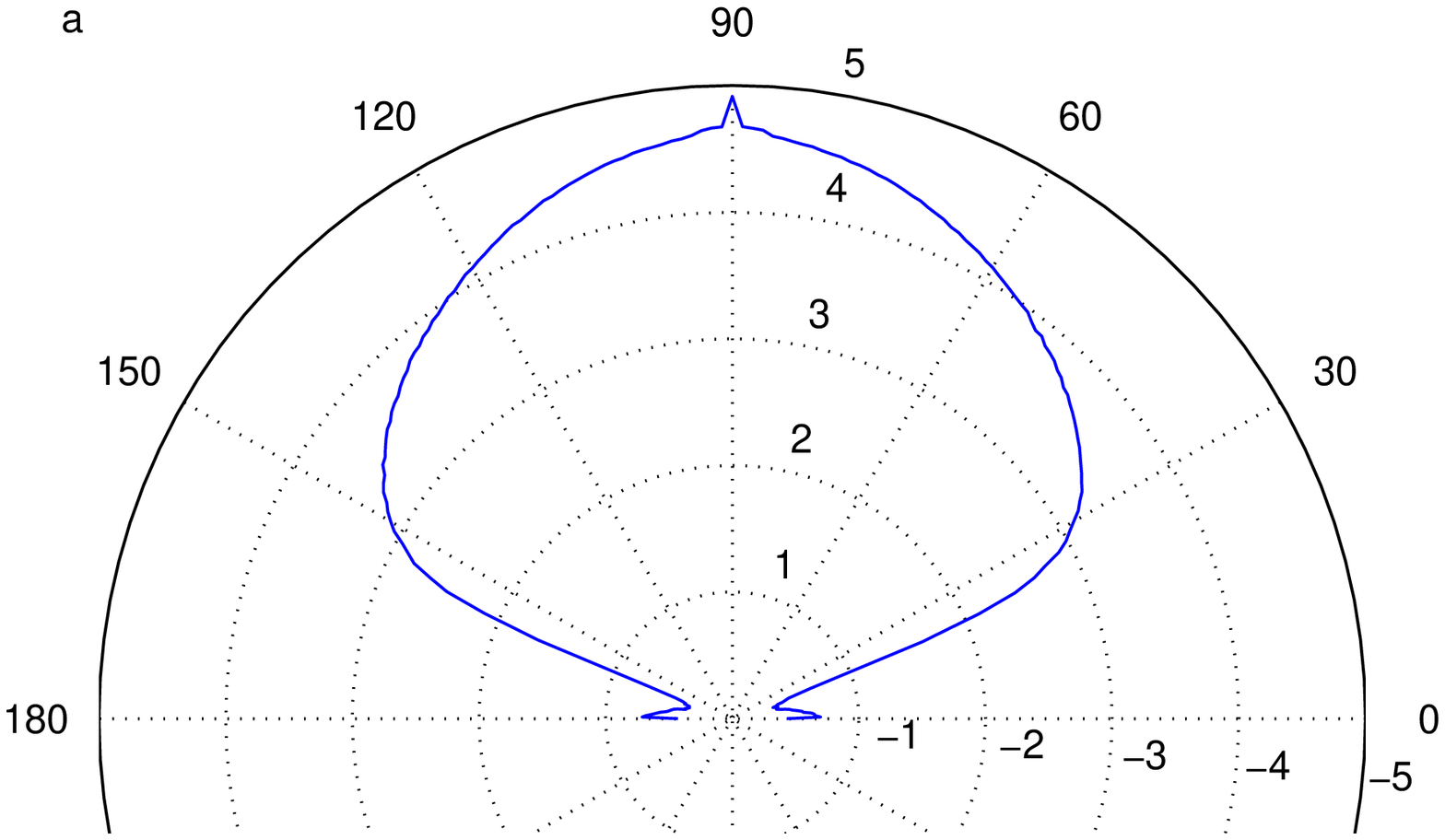,width=0.95\columnwidth,draft=false}  
\end{minipage}
\hfill
\begin{minipage}[t]{0.95\columnwidth}
\psfrag{b}[c][c]{(d)}
\epsfig{file=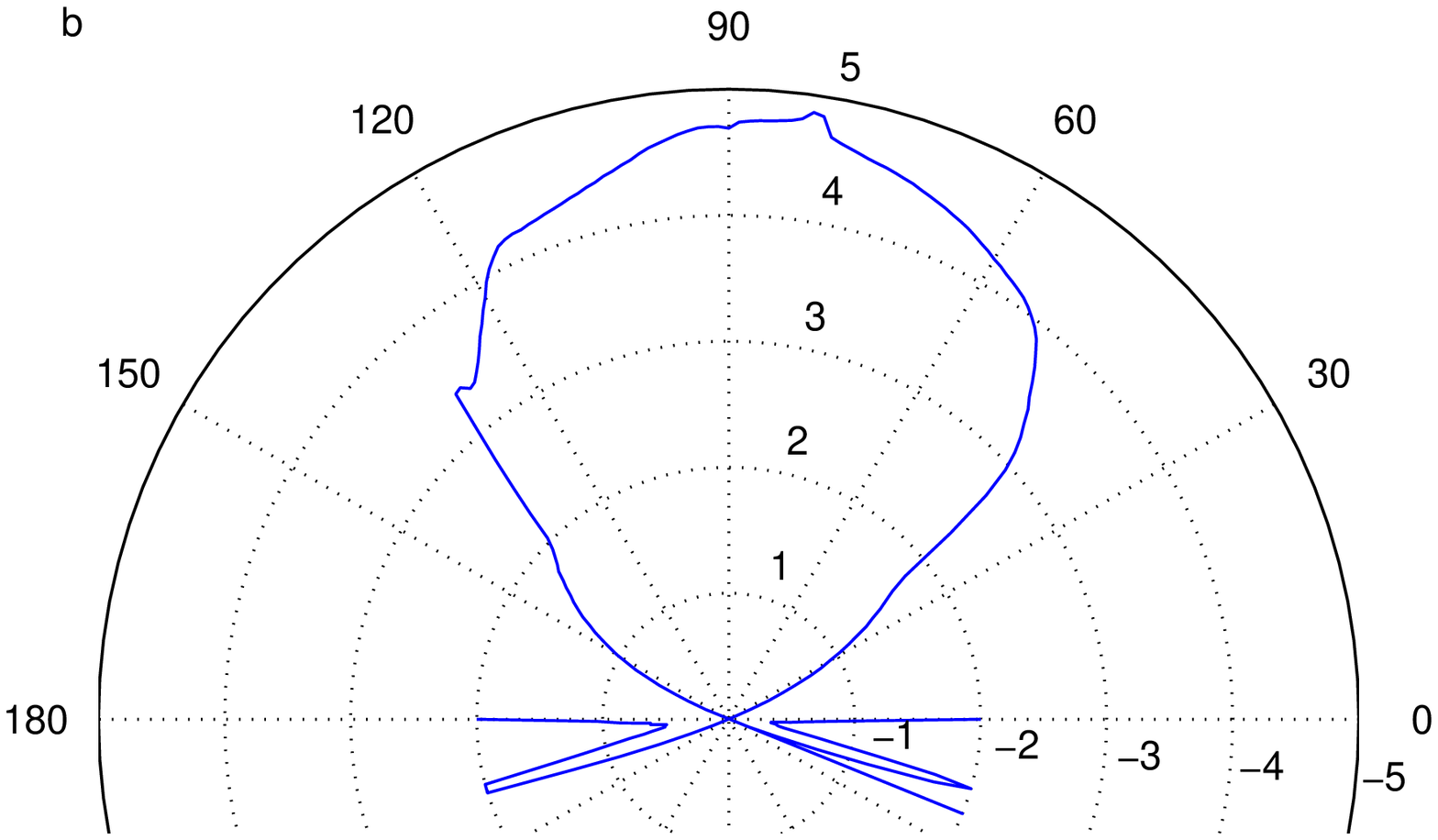,width=0.95\columnwidth,draft=false}  
\end{minipage}
\\[5pt]
\begin{minipage}[t]{0.95\columnwidth}
\psfrag{b}[c][c]{(e)}
\epsfig{file=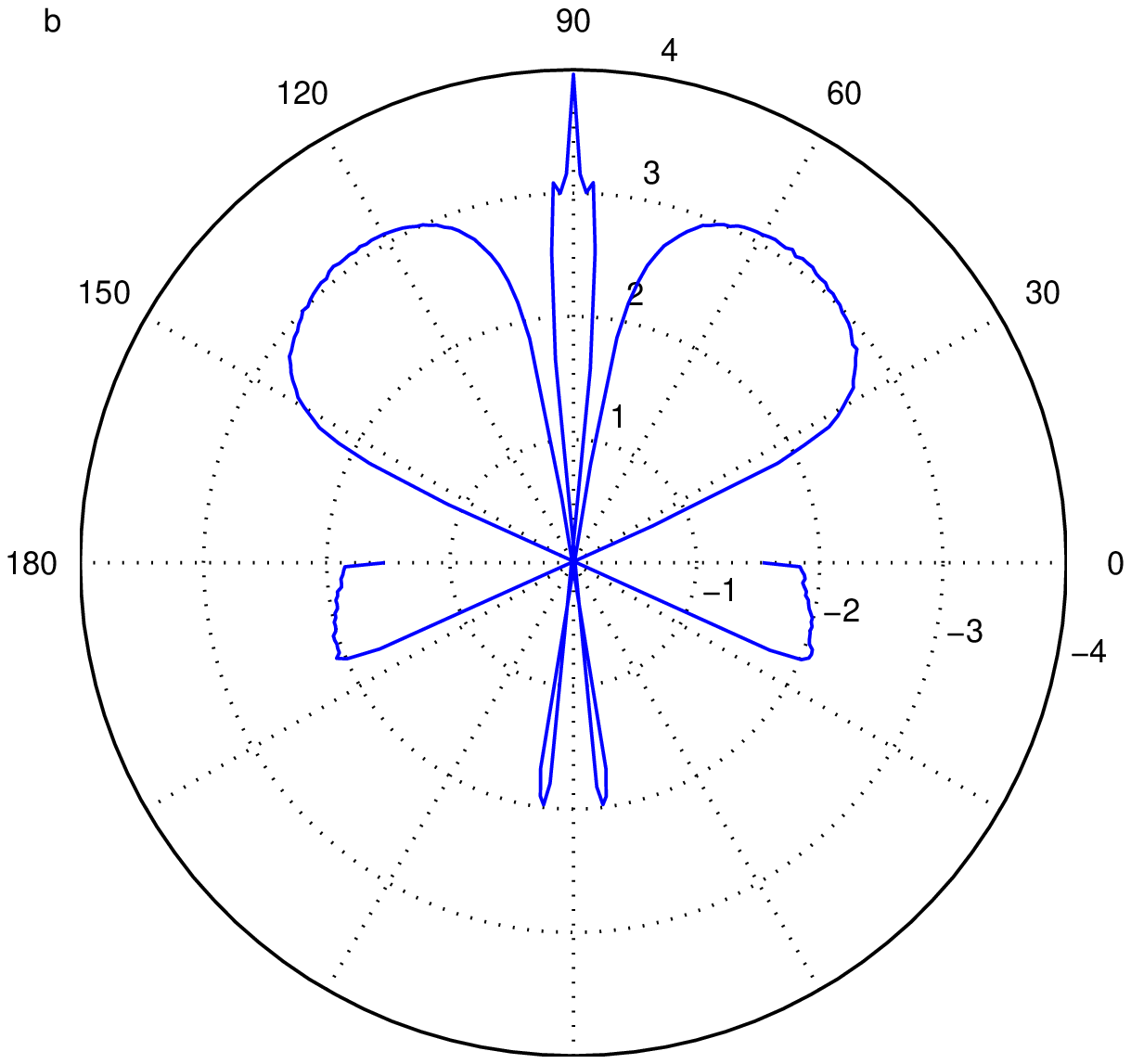,width=0.95\columnwidth,draft=false}  
\end{minipage}
\hfill
\begin{minipage}[t]{0.95\columnwidth}
\psfrag{b}[c][c]{(f)}
\epsfig{file=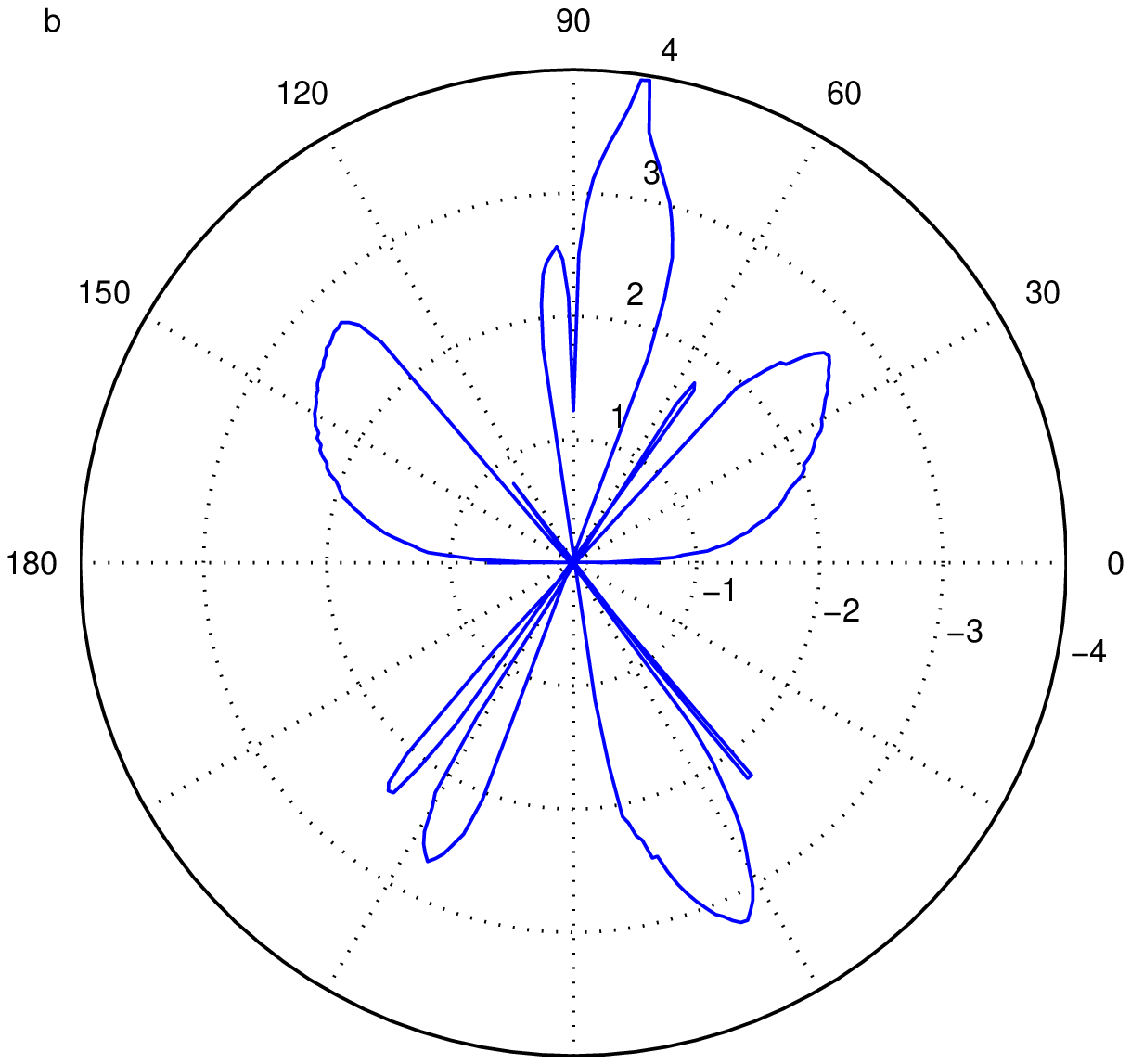,width=0.95\columnwidth,draft=false}  
\end{minipage}
\caption{ Polar plots for a 3\mm\ drop showing the intensity distribution along: (a)
$\Theta=82\de$ and (b) $\Phi=90\de$, both for unpolarised light; (c) $\Theta=82\de$ for
$y$-polarised light and (d) $\Phi=90\de$ for $z$-polarised light. The graphs in (e) and (f)
show the difference between the results for $y$- and $z$-polarised light along $\Theta=81\de$
and $\Phi=90\de$ respectively. Negative values are plotted into the lower half of the
circles.} \label{fig:pol2}
\end{figure*}

\fig{fig:cont2} shows the scattering patterns for a 3-mm drop and $10^8$ photons. In
\fig{fig:cont2}(a) the incoming light was unpolarized. The pattern exhibits a symmetry about
the central meridian that is to be expected from the rotational symmetry of the drops about
their vertical axis. This can also be observed in the corresponding polar plot of
\fig{fig:pol2}(a) that shows the intensity distribution along the $\Theta=82\de$ line that
contains the absolute maximum. The concentric semicircles are lines of constant intensity, and
the numbers give the corresponding intensities as powers of ten. \fig{fig:pol2}(b) shows the
distribution along the $\Phi=90\de$ meridian. A distinguishing feature is the lack of symmetry
about the equator. The absolute maximum has shifted from the equator ($\Theta=90\de$) to
$\Theta=82\de$. It is accompanied by another local maximum at $\Theta= 130\de$. For smaller
drop sizes, the absolute maximum shifts closer to the (90\de,90\de) direction again.
\par
Another prominent feature in \fig{fig:cont2}(a) is the sharp gradient, creating a steep
intensocline that separates the central area of relatively high intensity from the outer,
low-intensity regions. It is identified by the narrowly spaced contour lines. The model was
used to record the intensities from the rays $p=0$ and $p=1$ separately. The light that
arrives outside the intensocline has mostly undergone a single reflection at the first
interface, whereas the high-intensity area is mainly lit by the twice-refracted rays.
\par
\fig{fig:cont2}(b) shows the pattern for the same drop but with incident light polarized parallel to the $y$ axis.
There are two distinct minima near the equator at extreme angles for $\Phi$. This can also be seen on the corresponding
polar plot \fig{fig:pol2}(c). The sum of contributions from all angular bins is slightly ($\approx 1\%$) larger than
for unpolarized light. The opposite is true if the incident light is polarized parallel to the $z$ axis as shown in
\fig{fig:cont2}(c): The distinct minima are on the $\Phi=90\de$ meridian, symmetrically at $\Theta=90\de\pm 74\de$.
Light that is reflected into the angular bins containing the observed minima in Figs.~\ref{fig:pol2}(c) and
\ref{fig:pol2}(d) originates from points on the drop's surface where the angle of incidence is at the Brewster angle of
$\alpha_p=53.06\de$ and where the plane of polarization of the incident light is parallel to the scattering plane.
\par
\fig{fig:cont2}(d) shows the difference between the results obtained for $y$-polarized and $z$-polarized incident
light. This plot shows the absolute deviation, and the minima that can be seen in Figs.~\ref{fig:cont2}(b) and
\ref{fig:cont2}(c) do not appear. There is, however, a significant difference in the high-intensity area, inside the
intensocline and especially near the absolute maximum. Note that the scale of this last plot is nonlogarithmic to allow
for both positive and negative differences. A high-intensity ridge that extends from $\Theta=78\de$ to $\Theta=86\de$
on the central meridian contains values that exceed the maximum value covered by the scale (compare with
\fig{fig:pol2}(f)). The actual maximum in \fig{fig:cont2}(d) is 9200 and is located at $\Theta=81\de$. The maximum
difference of 9200 photons at the central maximum is equivalent to 12.5\% in relative terms, which explains why this
feature does not appear on the individual contour plots in Figs.~\ref{fig:cont2}(b) and \ref{fig:cont2}(c). If the
pattern in \fig{fig:cont2}(d) is divided by the (nonlogarithmic) intensity from \fig{fig:cont2}(a) (result not shown),
the extrema outside the intensocline visible in Figs.~\ref{fig:cont2}(b) and \ref{fig:cont2}(c) can be observed. The
intensity difference at these extrema is nearly twice the intensity value that is obtained there for unpolarized
incident light. If the incident light is unpolarized, the reflectance $R$ at $\alpha_p$ is half of the value of
$R_\bot$. Because $R_\|$ is equal to zero, the difference between the $y$- and $z$-polarized intensity is equal to
$R_\bot$. Dividing $R_\bot$ by $R$ for unpolarized light thus always yields 2. The finite size of the angular bins
allows inclusion of rays for which the angle of incidence differs slightly from $\alpha_p $. This should always keep
the absolute value of the extremes slightly below 2. A drop of radius $a=0.5\mm$, for example, has its extreme values
\begin{figure*}[t!]
\begin{center}
\epsfig{file=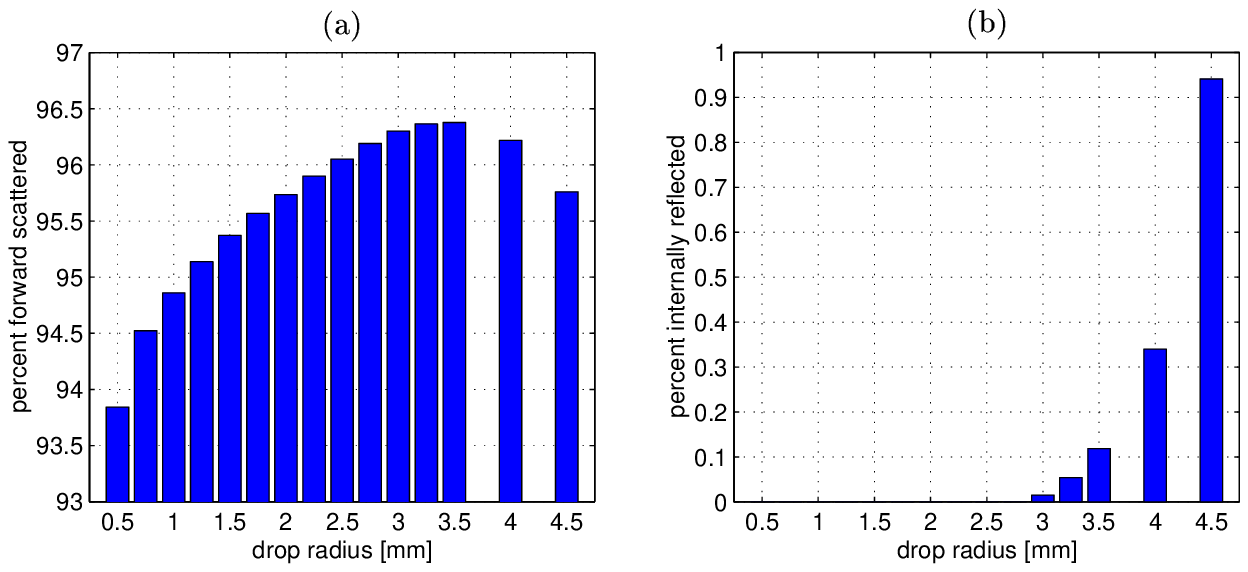,width=1.6\columnwidth}
\end{center}
\caption{(a) The fraction of the total incident intensity that is forward scattered and (b) the fraction of all
incident rays that are entirely internally reflected (regardless of their intensity).} \label{fig:2bars}
\end{figure*}
at 1.98 and -1.95, and the corresponding values for spherical drops are $\pm 1.95$. The reason why this is not observed
for larger drops along the central meridian might be a combination of the varying curvature along the vertical axis and
statistical fluctuations because the bins at such extreme angles contain only approximately 35 photons out of the total
of 100 million.
\par
\fig{fig:pol2}(e) gives a clearer image of what happens along the $\Theta=81\de$ latitude. The scale is again
logarithmic to accommodate the large spread of values. The pattern is dominated by the central maximum and two broad
maxima on each side. For smaller drop sizes the central maximum becomes less dominant and is almost nonexistent for a
1-mm drop. The two peripheral maxima, however, remain almost constant in size and location.
\par
\fig{fig:pol2}(f) shows the intensity distribution along the central meridian. It is
characterized by closely spaced maxima and minima. There is a general decrease in the number
of significant extrema to be observed with decreasing drop size, accompanied by a general
increase in overall symmetry.
\par
Over 99.5\% of the forward-scattered light from spherical scatterers emerges from either single reflection (ray $p=0$)
or twofold refraction ($p=1$). The assumption was made that this would be similar for distorted drops, thus justifying
rays with $p>1$ being neglected.
\par
\fig{fig:2bars}(a) shows the percentages of forward-scat-tered intensity in $p=0$ and $p=1$ for unpolarized incident
light and drop sizes ranging from $a=0.5\mm$ to $a=4.5\mm$. The fraction of forward-scattered intensity increases with
drop size, reaching a maximum for drops with $a=3.5\mm$. The rapid decline for larger drops can be explained from
\fig{fig:2bars}(b), which shows the percentage of photons that are internally reflected and thus discarded by the
model. It can be seen that internal reflection can be neglected for most drop sizes and becomes noticeable only for
large drops.
\par
To be able to give a quantitative comparison with some tabulated values, the model was used to find the scattering
pattern that would be produced by unpolarized light that is incident on a sphere of radius 0.1\mm. The result is given
in Table~\ref{tab:7}. It compares
\begin{table}[b!] \caption{Comparing results from the Computer Model and van de Hulst \cite{hulst}.} \label{tab:7}
\small
\begin{center}
\begin{tabular}{|l| c| c|}
\hline
 Source: & model & van de Hulst \\
\hline \hline
\multirow{2}{3.2cm}{\shortstack{\% forward scattered \\ in p=0 and p=1}}  & & \\
 &    \multirow{1}{1cm}[2.7mm]{93.73} & \multirow{1}{1cm}[2.7mm]{93.89}    \\
\hline
\end{tabular}
\end{center}
\end{table}
the fraction of energy that is scattered forward in rays $p=0$ and $p=1$ from the computer model and Table~\ref{tab:4}.
\begin{figure*}[t!]
\footnotesize
\begin{minipage}[t]{0.95\columnwidth}
\psfrag{a}[c][c]{(a)}
\epsfig{file=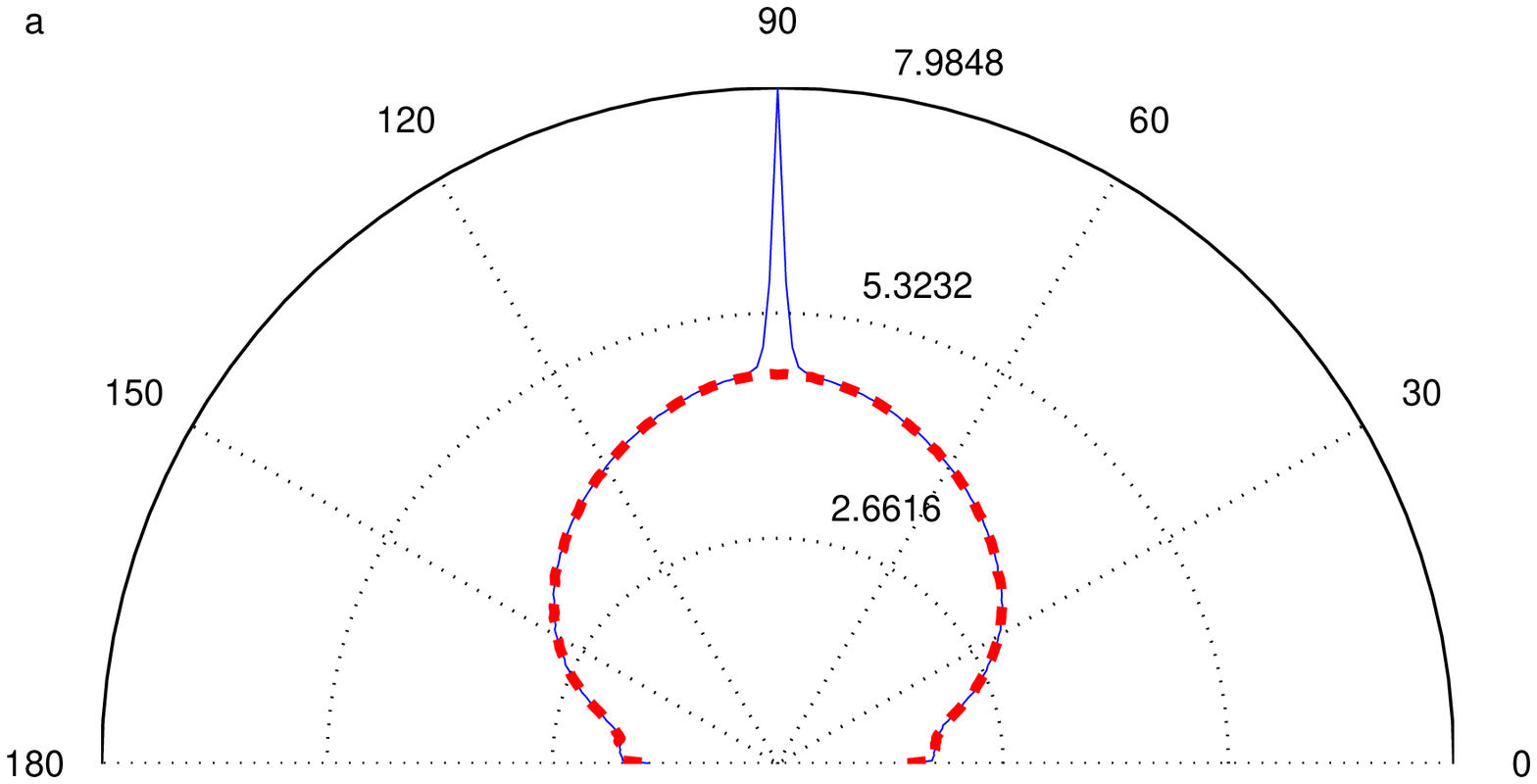,width=\columnwidth}  
\end{minipage}
\hfill
\begin{minipage}[t]{0.95\columnwidth}
\footnotesize \psfrag{a}[c][c]{(b)}
\epsfig{file=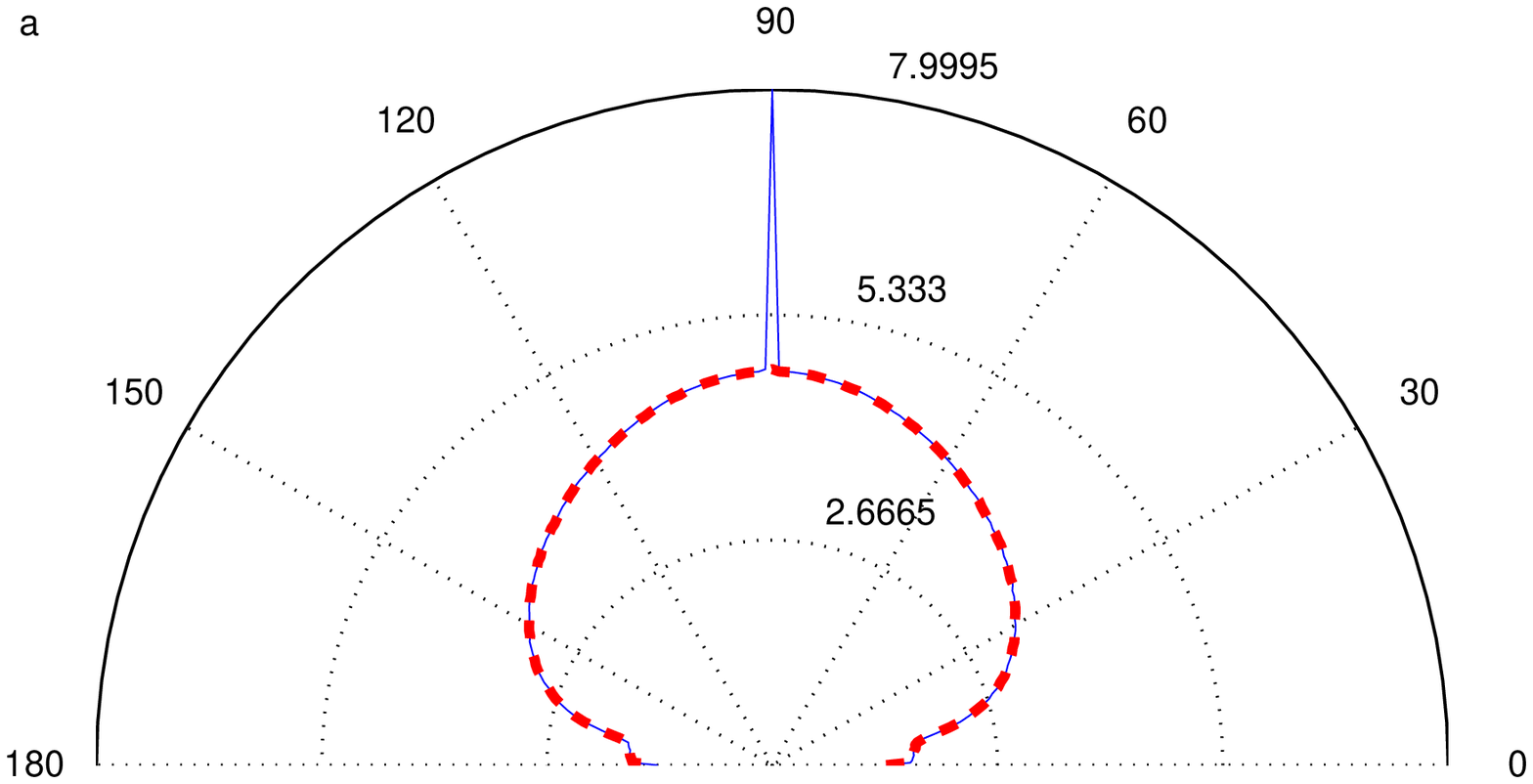,width=\columnwidth}  
\end{minipage}
\caption{Superposition of the polar plots of the logarithmic intensity including all components (continuous line) and
without diffraction (dashed line) along the 90\de\ latitude for (a) $a=0.1\mm$ and (b) $a=3\mm$.} \label{fig:bthu}
\end{figure*}
These differ by less than 0.2\%, justifying the model and its assumptions.
%
\subsection{Scattering from Single Drops -- including Diffraction} \label{sec:res2}
\begin{figure*}[tb!]
\footnotesize
\begin{minipage}[t]{0.95\columnwidth}
\psfrag{a}[cb][cb]{(a)} \psfrag{x}[ct][ct]{$\Phi$ [\de]} \psfrag{y}[ct][ct]{percentile change}
\epsfig{file=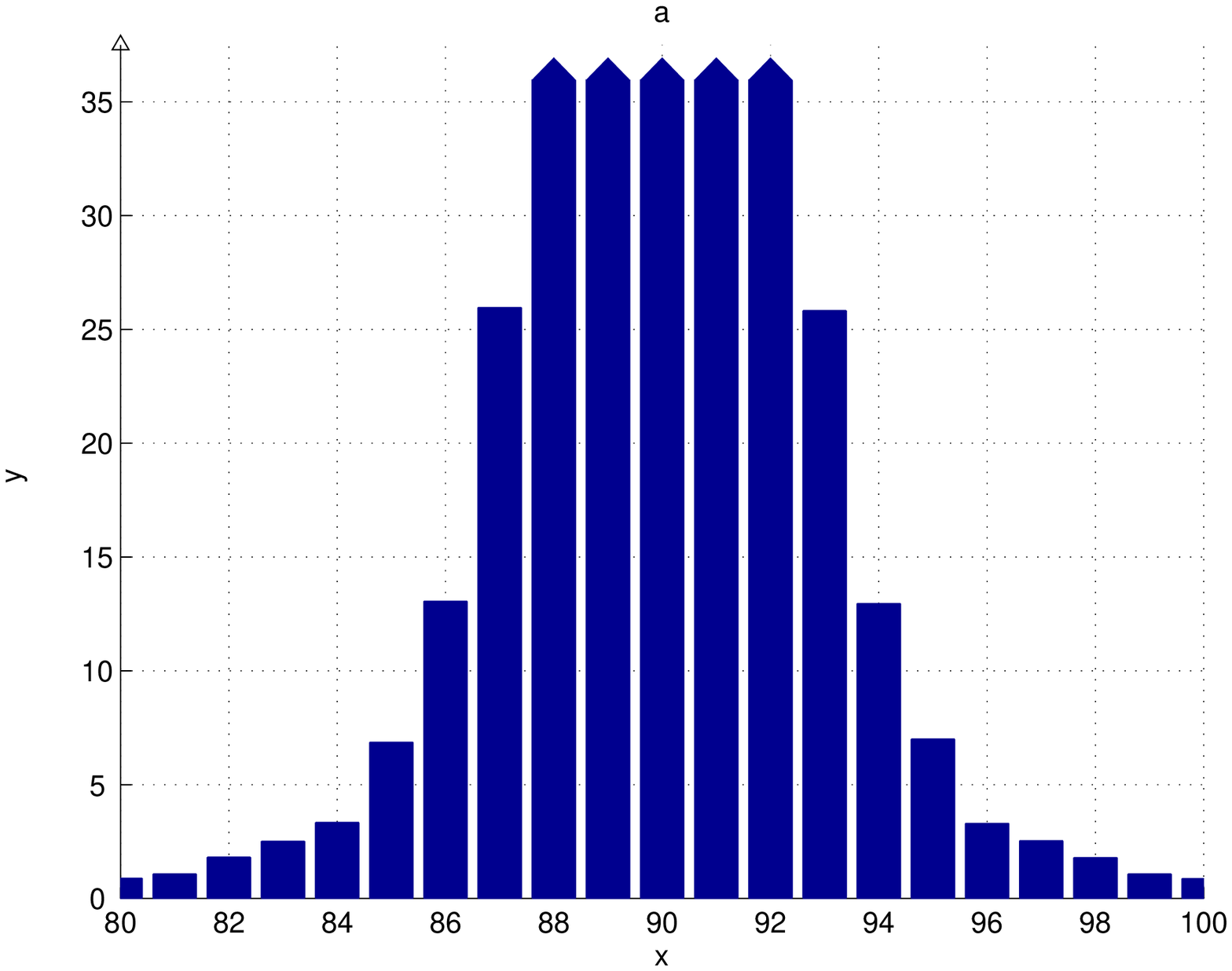,width=.85\columnwidth}  
\end{minipage}
\hfill
\begin{minipage}[t]{0.95\columnwidth}
\footnotesize \psfrag{a}[cb][cb]{(b)} \psfrag{x}[ct][ct]{$\Phi$ [\de]} \psfrag{y}[cb][ct]{percentile change}
\epsfig{file=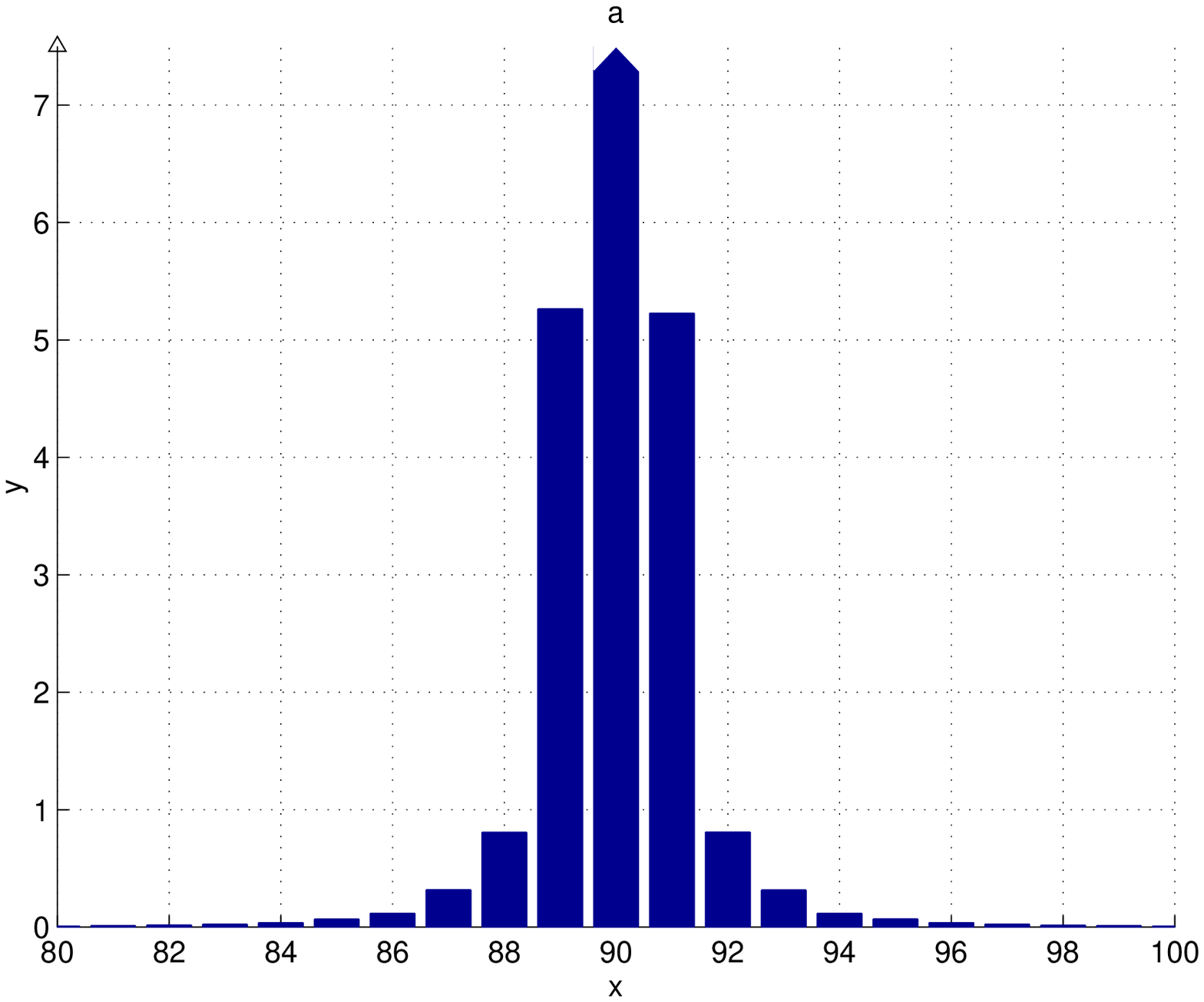,width=.85\columnwidth}  
\end{minipage}
\vspace*{2pt} \caption{The influence of diffraction on the overall scattering pattern. The percentile change between
the ray trace only and the ray trace plus diffraction is shown along the 90\de\ latitude for (a) $a=0.1\mm$ and (b)
$a=3\mm$.} \label{fig:chng}
\end{figure*}

%
The diffraction pattern is dominated by a narrow diffraction peak in the central forward direction. The width of this
diffraction peak depends on the size of the scatterer. The effects of diffraction are therefore demonstrated for two
extreme drop sizes: a 0.1-mm spherical drop and a 3-mm drop. Figures~\ref{fig:bthu}(a) and \ref{fig:bthu}(b) show the
combination of diffraction and ray trace together with the results from the ray trace alone. The central maximum for
the 3-mm drop is much sharper and slightly higher than for the smaller drop. The solid and dashed curves only differ
noticeably near the central diffraction peak, whereas the intensity at other angles is almost unchanged. This is
significant for the design of an instrument to measure drop sizes. If the detectors are placed well off the central
diffraction peak, the results can be simulated with contributions from the ray trace only. This is illustrated in a
more quantitative form in \fig{fig:chng}. It shows the percentile change of the intensity distribution from the ray
trace after diffraction is added. For the large 3-mm drop, the pattern is affected only by diffraction for angles of
$\Psi\lesssim 1\de$ if we accept a 1\% margin of error. For the smaller drop the same criterion would yield the
condition $\Psi\lesssim 10\de$.
%
\section{Practical Application}
%
If a large number of individual field measurements is averaged, the underlying pattern linking
the rainfall rate with drop diameter shows an exponential character as first proposed by
Marshall and Palmer.\cite{palmer} They suggested the following empirical relationship, the
so-called Marshall–Palmer distribution:
\begin{equation}
N(D)=N_0\exp{(-\Lambda D)}, \qquad N_0=0.08 \,{\rm cm}^{-4}; \label{M-P}
\end{equation}
where $N(D){\rm d}D$ is the number of drops per unit volume having diameters between $D$ and
$D+{\rm d}D$. The intercept parameter $N_0$ gives the intercept with the ordinate for $D=0$.
The distribution depends entirely on the parameter $\Lambda$ that is determined by the
rainfall rate $R$: $\Lambda=41R^{-0.21}$ where $R$ is in millimeters per hour and $\Lambda$ is
in inverse centimeters.
\par
\begin{figure}[b!]
\begin{center}
\epsfig{file=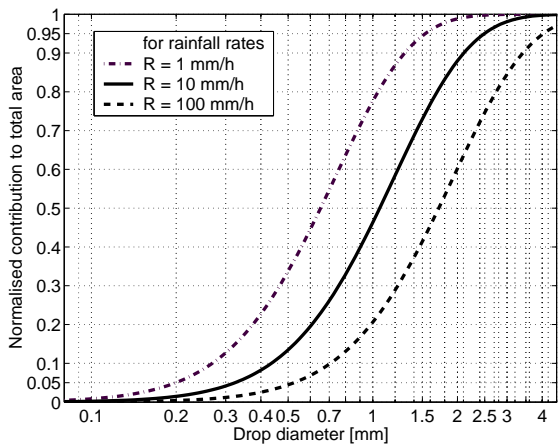,width=.9\columnwidth}%
\end{center}
\caption{Area presented to the incident light with drop diameter for three different rainfall
intensities.}\label{fig:areadist}
\end{figure}

We can now establish a function $P(D)$ to represent the cumulative probability of finding a
drop with a diameter between 0 and D in a large enough sample:
\begin{equation}
P(D)=1-\exp{(-\Lambda D)}    \label{eq:cumsum}
\end{equation}
If the drops are approximated as spheres of area $\pi/4\,D^2$ (which is possible because the
number of drops decreases rapidly as the distortion increases —- see \eq{eq:cumsum}) and by
integrating \eq{M-P} (while neglecting the constants $N_0$ and $\pi/4$ that both cancel out in
the normalization), one obtains the following area distribution:
\begin{eqnarray}
\integ_0^D {D\dash}^2 \; \exp{(-\Lambda D\dash)} \rm{d} D\dash = -\exp{(-\Lambda D)} \nonumber
\\
\times \left(\frac{D^2}{\Lambda}+\frac{2D}{\Lambda^2}+\frac{2}{\Lambda^3}\right)+
\frac{2}{\Lambda^3}\, . \label{eq:areadist}
\end{eqnarray}
Normalisation yields:
\begin{eqnarray}
\frac{\integ_0^D {D\dash}^2 \; \exp{(-\Lambda D\dash)} \rm{d} D\dash}
     {\integ_0^{\infty} {D}^2 \; \exp{(-\Lambda D)} \rm{d} D}=
-\exp{(-\Lambda D)} \nonumber \\
\times \left(\frac{\Lambda^2 D^2}{2}+D\Lambda+1\right)+ 1 \, . \label{eq:nareadist}
\end{eqnarray}
\fig{fig:areadist} shows the result for different rainfall intensities. For $R=10\,$mm/h, more than 50\% of the
incident light falls on drops with $D>1\mm$, which is where drops start showing first signs of distortion. For heavier
rain, more larger drops are present; and, as in the example of a $R=100\,$mm/h event, approximately 80\% of the
incident light falls on distorted drops. This suggests that drops with $D< 0.2\mm$ do not contribute significantly to
the scattered intensity and can be neglected, which would produce only a 5\% error even for light rain with
predominantly small drops.
%
\subsection{Multiple Scattering}
\begin{figure}[b!]
\begin{center}
\epsfig{file=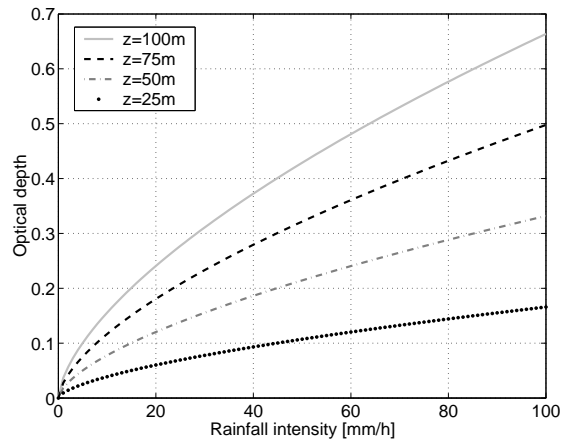,width=.9\columnwidth}%
\end{center}
\caption{Optical depth as a function of rainfall intensity for four different path lengths.} \label{fig:odepth}
\end{figure}

%
The optical depth of a path of length $z$ can be defined as:
\begin{eqnarray}
\tau_z=N_e\,C\,z\, , \label{eq:tau}
\end{eqnarray}
where $N_e$ is the number of drops per unit volume in the sample and $C$ is the extinction cross section. As was
discussed in Section 3, the extinction cross section approaches a limiting value of $2A$ (where $A$ is the area of the
scatterer) for particles with $x\gg 1$, and thus
\begin{eqnarray}
\tau_z=2\pi a^2 N_e\,z
\end{eqnarray}
where the drops have again been approximated as spheres. $N_e$ is given by
\begin{equation}
N_e=\integ _0^{\infty}N_0\exp{(-\Lambda D)}\d D \,. \label{eq:mpdint}
\end{equation}
Making the substitution, we find the area-distribution from \eq{eq:areadist} in a slightly
different form:
\begin{eqnarray}
\tau_z & = & 2\pi N_0 z \integ _0^{\infty} a^2 \exp{(-\Lambda D)} \d D \\
       & = & 4\pi N_0 z \integ _0^{\infty} a^2 \exp{(-2\Lambda a)} \d a \\
       & = & \frac{\pi N_0}{\Lambda^3}\,z \, .
\end{eqnarray}
\fig{fig:odepth} shows the optical depth as a function of rainfall intensity for four different path lengths. Van de
Hulst\cite{hulst} argues that single scattering prevails if $\tau_z < 0.1$; corrections for double scattering are
necessary for $0.1 < \tau_z < 0.3$, and multiple scattering is observed for even larger values of $\tau_z$. Other
sources\cite{hinkley} maintain that multiple scattering can be neglected for $\tau_z$ values of up to one.
%
\subsection{Numerical Simulation}
%
The rainfall event is simulated when a large volume $V$ is created above an imaginary laser beam and filled with drops
according to the Marshall–Palmer distribution for a given rainfall intensity. Only drops in the size range $0.2\mm \le
D \le 8\mm$ were considered for the
reasons outlined after \eq{eq:nareadist}. The total number of drops in $V$ is given by
\begin{equation}
N_V=V\,\integ_{0.02}^{0.8}N_0\exp{-\Lambda D}\d D\approx\frac{V\,N_0}{\Lambda}\exp{(-0.02\Lambda)} \label{eq:n}
\end{equation}
with the approximation $\exp{(-0.8\Lambda)}\approx 0$ yielding negligible errors even for high
rainfall intensities.
\par
It has been shown experimentally (e.g., Ref. \cite{joss}) that, for a constant mean rainfall
rate $R$, the number of drops is Poisson distributed around the mean. $N_V$ is therefore used
as a mean value to produce a Poisson deviate $N_P$. Once the volume is filled with $N_P$
drops, each drop is randomly assigned a size from an exponentially weighted distribution:
\begin{equation}
D=-\frac{1}{\Lambda}\ln{x} \: \: \mbox{ with } \: x\in\; ]0,1] \label{randi}
\end{equation}
where $x$ is a randomly generated number. The drops are then released and fall at their
terminal velocities (with the values from Gunn and Kinzer\cite{gunn}) through $V$. Every $t$
seconds a snapshot is taken, recording the number of drops present in the beam and their
sizes.
%
\subsection{Results}
%
The scattering patterns are obtained when the patterns from individual drops are combined
according to their abundance in the beam. The patterns from the individual drops are
normalized with the cross-sectional area to obtain a constant intensity per unit area. The
\begin{figure}[tb!]
\begin{center}
\epsfig{file=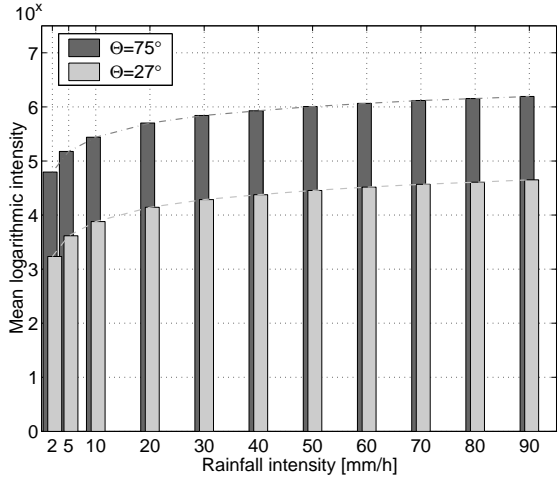,width=.9\columnwidth}%
\end{center}
\caption{Dependence of average received light intensity on rainfall rate $R$ for $(\Theta,\Phi)=(90\de,75\de)$ and
$(\Theta,\Phi)=(90\de,27\de)$.} \label{fig:means}
\end{figure}
\begin{figure}[tb!]
\begin{center}
\epsfig{file=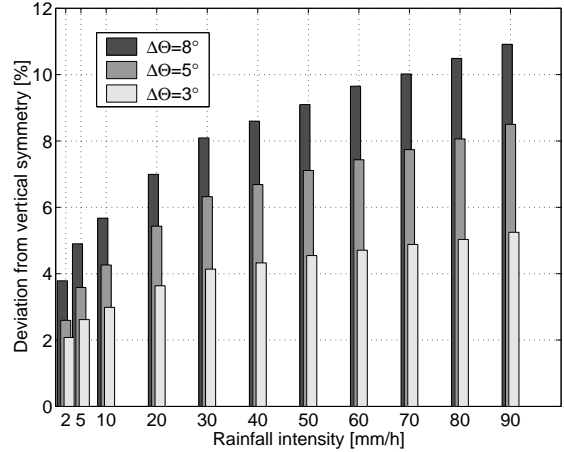,width=.9\columnwidth}%
\end{center}
\caption{Relative deviations from vertical symmetry in the scattering data for different
rainfall intensities and different angular bins. The deviation is given as a percentage of the
total received intensity at $(\Theta,\Phi)=(90\de$+$\Delta\Theta,90\de)$.}
\label{fig:vertasym}
\end{figure}
result is a time series that gives the variations of the received intensity for each angular
bin.
\par
As expected, the results show that the higher the rainfall rate, the higher the mean scattered
intensity over the sampling period. This is shown in \fig{fig:means} for two different angular
bins. The mean detected intensity is thus a conclusive indication of the rainfall intensity
$R$. The degree of asymmetry about the equator can also be used to infer the rainfall rate.
\fig{fig:vertasym} shows the difference between angular bins at $(\Theta,\Phi)=
(90\de$--$\Delta\Theta,90\de)$ and $(\Theta,\Phi)=(90\de$+$\Delta\Theta,90\de)$. As can be
seen, the amount of light received in the bin below the equator is considerably higher than
for the corresponding element symmetrically above the equator. The difference increases as
expected with rainfall intensity and varies depending on the distance to the equator.
\par
Although the light intensity is a stable indicator of the rainfall rate, the effects of wind
on the orientation of the vertical axis of the drop can be severe on the above symmetry
considerations. Other factors that are likely to occur in real situations and that are
neglected in this model are coalescence, drop breakup, and oscillations, as well as
multiple-scattering events. Possible realizations of field experiments are given in Ref.
\cite{ross}.
%
\section{Summary} \label{sec:sum}
%
The distorted shape of raindrops at terminal velocity is often ignored and approximated with
either spheres or ellipsoids. In contrast, the numerical model developed in this study
determines the scattering pattern for the true drop shapes. This was achieved through a
combination of geometrical optics and a statistical Monte Carlo technique. The ray trace
yielded scattering patterns for different drop sizes and polarizations of the incident light.
The shape of the drops could be inferred from the varying degrees of depolarization of the
scattered light and the asymmetries observed in the overall scattering behavior. No
experimental data are available for light scattering from distorted drops. However, the
results obtained for small (spherical) drops match tabulated values confirming the validity of
the model.
\par
We have treated the problem of diffraction by successfully approximating the drops as
ellipsoids. The error incurred has been found to be negligible even for the largest drops
examined. Adjustments for the results from the ray trace were necessary only near the central
diffraction peak whereas diffraction can be neglected for the remaining pattern.
\par
The asymmetries in the scattering behavior can be used for rainfall measurements with a laser
and several detectors as the drop size distribution depends on the rainfall intensity and thus
influences the overall symmetry. Further details including the experimental realization and
application to high-resolution measurements of rainfall structures can be found in
Ross.\cite{ross}.
\par
This model was developed as part of a Diplomarbeit (equivalent to a Master of Science project)
at the Fachbereich Physik of the Freie Universit\"at Berlin conducted in close collaboration
with the Physics Department of the University of Auckland that generously provided the
resources for this research.

\end{document}